\documentclass[
  reprint,
  amsmath,amssymb,
  aip,
  jap,       % Journal of Applied Physics
  floatfix,
  nofootinbib
]{revtex4-2}

\usepackage{amsfonts}      % if you use \mathbb
\usepackage{mathtools}     % extensions to amsmath
\usepackage{graphicx}      % figures
\usepackage{dcolumn}       % decimal-aligned columns in tables
\usepackage{bm}            % bold math
\usepackage{subcaption}    % subfigures (loads caption internally)
\usepackage{tensor}        % tensor indices
\usepackage{enumitem}      % list control
\usepackage[usenames,dvipsnames]{color}

\usepackage{float} % for [H] float placement

\usepackage{hyperref}
\hypersetup{
    colorlinks=true,
    linkcolor=blue,
    filecolor=magenta,
    urlcolor=cyan,
    pdftitle={How it cools? Studying the heat flow out of a semi-infinite slab in welding: An analytical approach},
    pdfpagemode=FullScreen,
}
\begin{document}

\title{How it cools? Studying the heat flow out of a semi-infinite slab in welding: An analytical approach}

\author{Fawzi Aly}
\email{maly@ewi.org}
\affiliation{Physics Department, SUNY at Buffalo, Buffalo, New York, USA}
\affiliation{EWI, Data Science Team, Buffalo, New York, USA}

\author{Alex Kitt}
\email{akitt@ewi.org}
\affiliation{EWI, Data Science Team, Buffalo, New York, USA}

\author{Luke Mohr}
\email{lmohr@ewi.org}
\affiliation{EWI, Data Science Team, Buffalo, New York, USA}

\begin{abstract}
Additive manufacturing and welding processes are highly sensitive to heat dissipation, where improper thermal management leads to residual stresses, distortions, and cracking. Existing heat transfer models, such as Rosenthal’s solutions, fail to handle finite 3D geometries, cooling effects, or transient behavior, limiting their accuracy. We overcome these limitations by developing an analytical framework that incorporates cooling boundary conditions mimicking Newton’s Law of Cooling. Using two different and proven-equivalent approaches—Laplace transform and Fourier series—we derive closed-form solutions for transient and steady-state temperature profiles under various heat sources, including Gaussian, ellipsoidal, double-ellipsoidal, and time-dependent on/off switch sources. We compare our analytical solutions to numerical implementations, demonstrating strong agreement while providing deeper physical insight. This approach significantly reduces computational cost and experimental requirements, making it a scalable tool for optimizing thermal predictions and mitigating residual stresses in metal-based manufacturing. Additionally, our framework enables the generation of synthetic datasets for machine learning models to predict heat distribution efficiently.
\end{abstract}

\maketitle

\section{Introduction}\label{Introduction}

Additive Manufacturing (AM) excels in applications requiring complex geometries, rapid prototyping, and low-volume production without the need for costly tooling. Its ability to create lightweight, optimized structures makes it ideal for aerospace, medical implants, and specialized industrial components \cite{Gibson2021,Ngo2018}. However, for AM to truly compete with traditional methods, advancements in material properties and defect predictability are essential \cite{DebRoy2018,Grasso2017,Lewandowski2016}.

In AM, the laser moves methodically, melting the powdered material and fusing it into a solid structure. The part appears flawless as it takes shape—but beneath the surface, \textit{something unseen is happening}. As the material cools, uneven heat dissipation creates thermal gradients, leading to internal stresses; without proper thermal management, microscopic cracks begin to form, weakening the structural integrity of the part \cite{DebRoy2018,Mercelis2006,Lewandowski2016,Khairallah2016}.

 Effective cooling management is essential to minimize residual stresses, microstructural changes, and crack formation, ultimately improving joint quality and mechanical performance \cite{Mercelis2006,Vrancken2016,Patterson2019,Parry2019,Khairallah2016,Ganeriwala2014,Hussein2013}. Nevertheless, the main question remains: \textit{Can we develop more accurate physics models to predict how a material cools?} Accurately capturing heat dissipation in AM and welding remains a challenge, requiring a combination of sophisticated analytical and numerical approaches. This study contributes to this ongoing effort by proposing a more refined framework for modeling heat dissipation.

The evolution of the system temperature is governed by the diffusion-convection equation, which serves as the fundamental model of heat transfer in this study. To accurately capture cooling effects, one must account for both the energy input from a moving heat source and the heat loss to the environment. However, this process is highly complex due to factors such as the material’s geometry, the non-uniform spatial distribution of the heat source, and its power variation as function of time \cite{rosenthal1946,eager1983,nguyen1999}.  

Traditional analytical approaches, as demonstrated in the literature by Rosenthal and others \cite{rosenthal1946,eager1983,nguyen1999}, often assume an open-domain system to take advantage of Fourier integral transformations. This method simplifies the problem by reducing it to a convolution of the heat source with the heat kernel over time. While effective for simple cases, these models fail to capture realistic cooling effects due to their inability to incorporate finite-width constraints and evolving source distributions. As a result, many previous models have struggled to accurately represent cooling, often oversimplifying heat dissipation and its impact on temperature evolution \cite{Kou2003,Goldak2005,Dowden2001,Pavelic1969,Michaleris2014}.

While numerical approaches, such as the finite-element method (FEM), can effectively handle some of these complexities, they are computationally expensive and often fail to provide clear physical intuition about the underlying processes \cite{Michaleris2014,Frazier2014,Khairallah2016}. Analytical approaches, on the other hand, offer a more efficient alternative, not only reducing computational cost and experimental effort but also providing deeper physical insight into heat transfer mechanisms \cite{carslaw1959conduction,Michaleris2014,rosenthal1946}. Additionally, they can be seamlessly integrated into physics-informed machine learning models, further enhancing predictive capabilities in AM and thermal analysis \cite{Raissi2019,Karniadakis2021}.

In this article, we address the challenges of solving the heat equation under more realistic boundary conditions (B.C.). We also account for the transient behavior of system's temperature profile by employing the Laplace transformation (LT), often rendering the equation more amenable to separation of variables, and naturally incorporates B.C. and I.C. \cite{schiff1999laplace,krantz2022differential}. This approach is not merely an alternative analytical method but introduces more physically realistic conditions, such as  cooling effects and finite-width constraints, which are often absent in standard open-domain formulations. The resulting system of equations remains analytically tractable while capturing essential thermal behaviors. Moreover, these refined equations can still be implemented numerically if needed, providing a more physically accurate alternative to conventional numerical solutions in literature. While this method is particularly well-suited for initial-boundary value problems (IBVPs), the inversion from spectral space \( s \) back to the real-time domain \( t \) remains a challenge, often involving complex contour integrals with branch cuts. Nevertheless, we manage to overcome this obstacle and obtain a closed-form series solutions for the heat kernel, as well as for specific heat sources such as Gaussian and ellipsoidal spatial profiles and other time-dependent variations.  These series solutions are shown to be equivalent to Fourier series (FS) expansions, constrained by the initial data. Since FS are particularly effective for handling finite-domain B.C., whereas LT are more suited for IVPs, we propose a more rigorous approach that begins with LT and later establishes its equivalence to FS expansions. This dual formulation provides greater flexibility in implementing the final solution, as we will demonstrate in this work.

\begin{figure}
\includegraphics[width=8cm]{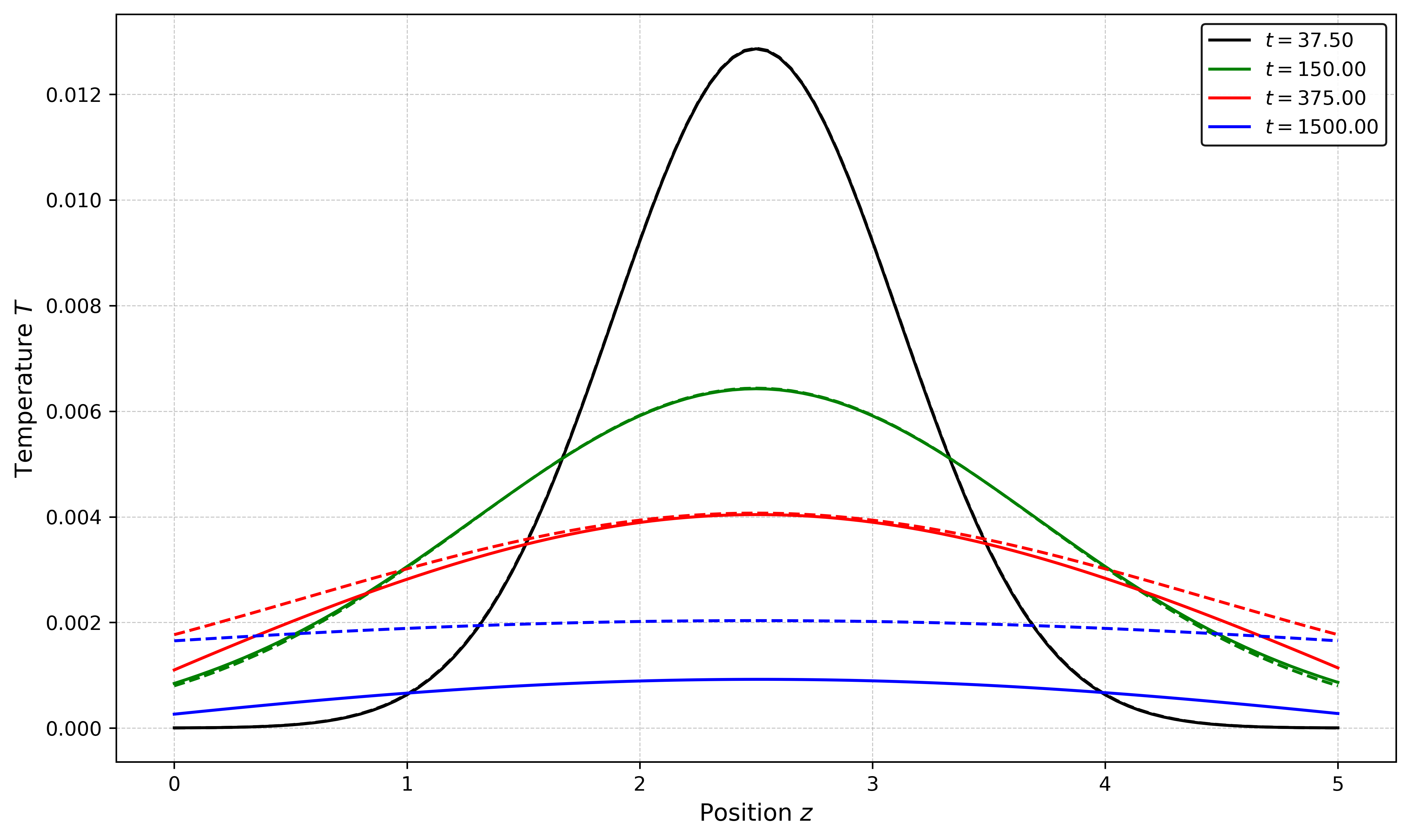}
\caption{Comparison between heat profile evolution for open-domain and finite-width solutions with cooling. The dashed lines represent the open-domain solution obtained via Fourier integral methods, while the solid lines correspond to the finite-width solution with NLOC with parameters: $H_1=1.2$ and $H_2=1.5$. Initially, both solutions coincide, indicating that heat didn't diffuse to the material boundaries yet. This overlap validates that, at early timescales, boundary cooling effects are not on yet. However, as time progresses to a critical point, characterized by the material width $w$, the source location $z_0$, and the thermal diffusivity $\alpha$, heat begins to reach the boundaries, activating cooling effects. Consequently, a noticeable difference between the two solutions develops, with substantial deviations at the boundaries over time, as shown in the red and blue profiles. This illustrates the significant role of boundary cooling in finite domains at later times of evolution.}
\label{fig:open_vs_finite}
\end{figure}

\subsection{Executive Summary}

This work develops an analytical framework for solving the three-dimensional heat transfer problem in finite geometries while incorporating cooling effects. Heat dissipation is modeled as an energy flux boundary condition, allowing for a more realistic description of heat transfer. The framework accounts for Newton's Law of Cooling (NLOC), Stefan-Boltzmann (SB) radiation, and a generalized phenomenological cooling model \cite{incropera2011fundamentals,siegel2015thermal}

The main contributions of this work are:
\begin{enumerate}
    \item \textbf{Modeling Cooling Effects:} We model the heat flux out of the surfaces of the finite-width slab using B.C. governed by NLOC. This framework is readily generalizable to incorporate both radiative heat loss and more complex phenomenological cooling behaviors.
     \item \textbf{Time-dependent heat sources:} We derive closed-form analytical solutions for pulsed and continuous sources, solving the on/off switch point source model in one dimension for moving sources and in two and three dimensions for static sources. A generalization to moving sources in three dimensions results in a one-dimensional integral suitable for numerical evaluation.
    \item \textbf{Spatially distributed sources:} The methodology is extended to Gaussian, ellipsoidal, and double-ellipsoidal profiles, generalizing previous models such as those by Nguyen and Eager \& Tsai. For short-duration sources, exact analytical solutions are obtained using two mathematically equivalent approaches:
    \begin{itemize}
        \item Two Fourier transforms (FT) and a LT.
        \item Two Fourier transforms and a FS.
    \end{itemize}
    Their equivalence is proven, and for long-duration sources, a one-dimensional integral formulation remains computationally efficient.
    \item \textbf{Heat kernel solution:} The three-plus-one-dimensional heat kernel is derived in closed form using both the LT and FS approaches and validated against three solution techniques:
    \begin{itemize}
        \item A fully numerical implementation of the model.
        \item The residue theorem.
        \item The FS method.
    \end{itemize}
    While all three yield equivalent results, analytical solutions significantly reduce computational costs and provide deeper insight into heat transfer dynamics.
    \item \textbf{Rosenthal model recovery:} The framework naturally recovers Rosenthal’s solutions for two-dimensional heat transfer with cooling and three-dimensional heat transfer without cooling as limiting cases, demonstrating its robustness and broad applicability.
\end{enumerate}

This work establishes a rigorous foundation for modeling heat transfer in finite domains, offering advancements in computational heat transfer and process optimization for manufacturing applications.

This paper is structured as follows: Section \ref{Introduction} provides an overview, followed by the problem definition and setup in Section \ref{Problem Statement}. Section \ref{How to Model Cooling Mathematically?} develops the mathematical framework, deriving the three-plus-one-dimensional heat kernel via integral transforms (Sections \ref{Integral Transform Approach for the 3+1 Heat Kernel} and \ref{Cooling Boundary Conditions}), analyzing singularities (Sections \ref{Commuting the integrals}, \ref{Poles and Singular structure}, and \ref{Branch Cut}), and introducing an alternative FS method (Section \ref{Alternative Approach: Fourier Series}). Section \ref{Heat Profile for different Source terms} examines heat distributions for Gaussian, ellipsoidal, and time-dependent sources (Section \ref{source Using LT}), while Section \ref{Rosenthal_as_special_case} recovers Rosenthal’s models as limiting cases (Sections \ref{Rosenthal 2D} and \ref{Rosenthal 3D}) and explores transient effects (Sections \ref{Transient Behavior}, \ref{SpecialCase LT Rosenthal}, and \ref{TransientOurModel}). Section \ref{Conclusion} summarizes key findings and future directions. The appendices provide detailed derivations, including point source modeling (Section \ref{source Using LT}), integral transformations (Section \ref{FS and LT}), initial condition effects (Section \ref{IC Section}), and more details on the derivation of our Green’s function (Sections \ref{Derive GF} and \ref{Formulas}).

\begin{figure}
\includegraphics[width=8cm]{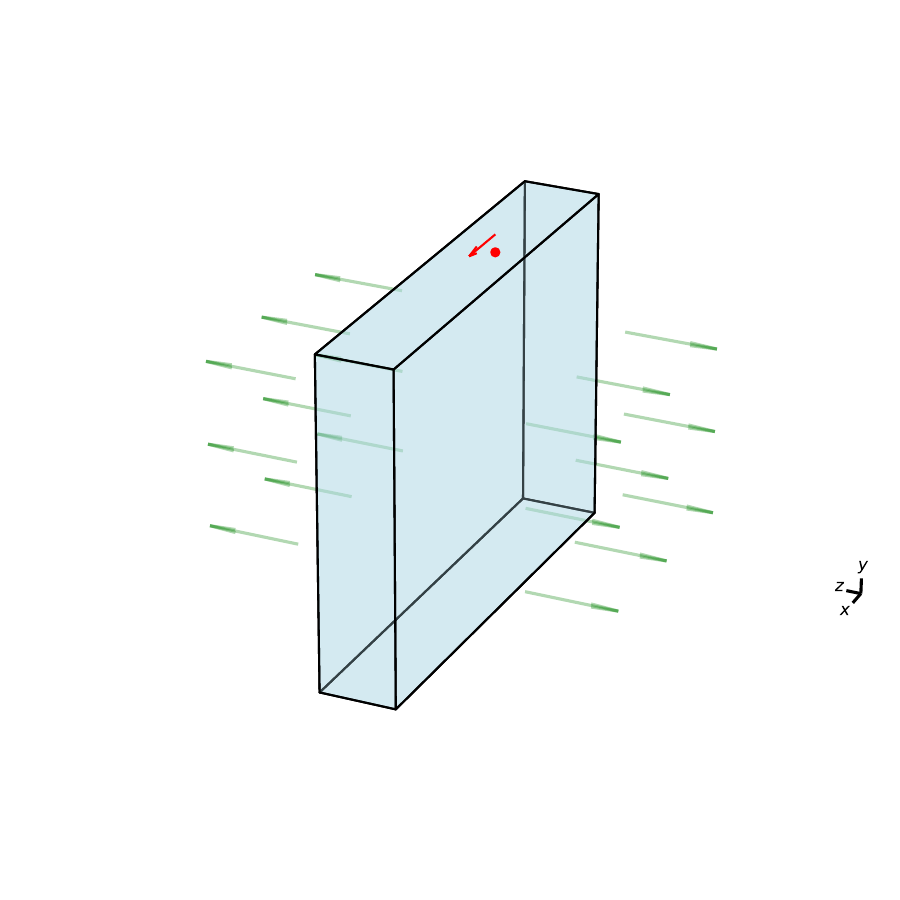}
\caption{This figure illustrates a semi-infinite slab (gray) with a heat source (red sphere) positioned at \( (\tilde{x}, \tilde{y}, \tilde{z}) = (0.4, 0.3, 0.2) \), indicating the location of interest for solving the heat kernel \( G \). Blue arrows represent the heat flux moving outward from both the top and bottom surfaces, modeled using NLOC as B.C. A red arrow above the sphere shows its motion in the positive \( \tilde{x} \)-direction, emphasizing the movement of the heat source across the slab—hence, this snapshot is taken in the lab frame $\vec{\tilde{x}}$. The black coordinate system \( (\tilde{x}, \tilde{y}, \tilde{z}) \), placed beside the slab, provides spatial reference and that \( \tilde{z} \) represents the finite-width $w=0.4$ dimension.}
\label{Sketch}
\end{figure}									

\section{Problem Statement}\label{Problem Statement}

\textit{Why and when is accounting for cooling crucial?} Consider a semi-infinite material subjected to a short pulse heating, such as the sample illustrated in Fig.~\ref{Sketch}. To model this process, we adopt two approaches:  

\begin{itemize}
    \item \textbf{Model A:} The \textbf{open-domain solution}, exemplified by the Rosenthal Model in 3D and similar models developed in prior works \cite{eager1983, rosenthal1946, nguyen1999}.  
    \item \textbf{Model B:} Our \textbf{finite-width model}, incorporating cooling B.C, as developed in this work.  
\end{itemize}

The corresponding temperature profiles at various time instances are compared in Fig.~\ref{fig:open_vs_finite}.  

Initially, heat propagates rapidly within the material, and the temperature distributions predicted by both models coincide, as cooling effects have yet to influence the system. Assuming the heat source behaves approximately as a point source, thermal diffusion occurs isotropically at a characteristic speed dictated by the thermal diffusivity \( \alpha \), which is assumed to be spatially uniform.  

However, after a critical time—when heat has reached the two faces of the sample—cooling effects encoded in the B.C. start playing a role. At this stage, Model A and Model B diverge, as the former neglects cooling. Over time, these effects penetrate deeper into the material, significantly altering the thermal profile in Model B. The reader can track this transition by comparing the \textit{dashed} (Model A) and \textit{solid} (Model B) curves at different moments: \( t = 37.5, 150, 375, 1500 \).  

For processes involving prolonged heating—such as in welding—the disparity between Model A and Model B becomes even more pronounced as the system approaches a steady state. This steady-state behavior is crucial for refining heat management algorithms and ensuring accurate thermal predictions. Thus, incorporating cooling effects is essential for realistic modeling of temperature evolution.  

\subsection{Setup}\label{Setup}

To systematically analyze this problem, we consider a semi-infinite slab subjected to a moving heat source traveling at a constant velocity \( v \) along the \( \tilde{x} \)-direction. The slab extends infinitely in \( \tilde{x} \) and \( \tilde{y} \) but has a finite width \( w \) in \( \tilde{z} \). Cooling effects, governed by Newton’s Law of Cooling (NLOC), are imposed as B.C. at both ends in the \( \tilde{z} \)-direction. Additionally, Stefan-Boltzmann (SB) radiation terms or polynomial homogeneous cooling effects can be incorporated for phenomenological applications (see Appendix \ref{Cooling Boundary Conditions} for details).  

\begin{figure}
    \centering
    \includegraphics[width=8cm]{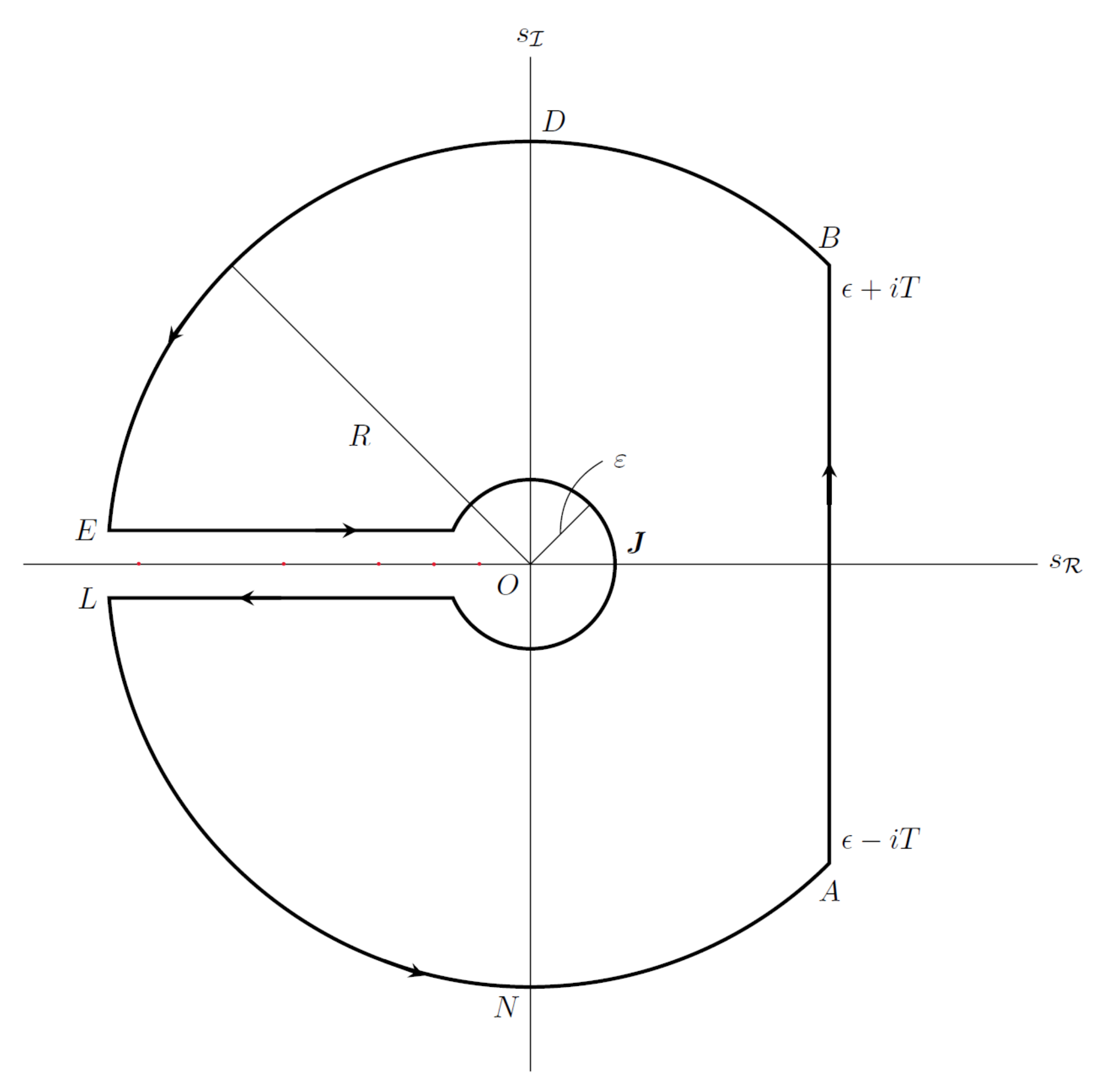}
    \caption{A schematic of the Bromwich contour integral used in the inverse LT. The contour, which includes a large arc defined by points $BDE$ and $ANL$, contributes zero to the integral by applying the Big Lemma theorem. A branch cut extends from \([0, \infty)\) along the real axis, and an infinite countable number of simple poles are located on the negative \( s_R \)-axis, evaluated using the residue theorem. Notably, there is no pole at \( s=0 \), and no contributions arise from the segments $EO$ and $LO$. This setup allows for capturing the essential behavior of the inverse under the conditions specified. The poles $s_n$ marked in red dots are on the negative $s_R$-axis, and the spacing between them increases monotonically with $n$.}
    \label{LT contour}
\end{figure}

\section{How to Model Cooling Mathematically?}\label{How to Model Cooling Mathematically?}

The heat transfer physics is described by the hyperbolic diffusion-convection equation, which can, in general, include nonlinear effects such as temperature-dependent properties, phase changes, radiative transfer, or non-Fourier conduction \cite{cattaneo1948,vernotte1958,carslaw1959conduction}. However, in this work, we focus on the linear system. The hyperbolic form of the heat equation in the lab frame $\vec{\tilde{x}}$ is given by:

\begin{equation}
\frac{\partial \tilde{T}}{\partial t} - \alpha \tilde{\nabla}^2 \tilde{T} = \tilde{S}(t, \vec{\tilde{x}}),
\end{equation}

where \( \tilde{T} = \tilde{T}(t, \vec{\tilde{x}}) \) represents the temperature distribution and $\tilde{S}(t, \vec{\tilde{x}})$ represents the heat source term, which generally depends on position $\vec{\tilde{x}}$ and time $t$.  

We perform the variable transformation \( \tilde{x} \rightarrow x - vt \), where \( v \) is the velocity of the moving source, transforming the problem into the source's co-moving frame $\vec{x}$. In this frame, the PDE takes the form:

\begin{equation}\label{equation in co moving frame}
\frac{\partial T}{\partial t} - v\frac{\partial T}{\partial x} - \alpha \nabla^2 T = S(t, \vec{x}).
\end{equation}

In the co-moving frame, the source term simplifies significantly: it reduces to a \((3+1)\)-dimensional Dirac delta pulse at the origin, which facilitates obtaining the Green's function \( G \). After that we can directly convolve it with the actual source term \( S \) to evaluate \( T \), see equation \eqref{Convlution}. Please note that we use \(G\) interchangeably with the terms \textit{heat kernel} and \textit{Green's function}.  

The approach employed in this work involves reducing the full \( 3+1 \) dimensional PDE to an effective one-dimensional ODE, which we then solve analytically when possible. This reduction is achieved through three integral transformations: two Fourier transformations (FT) in the spatial dimensions \( x \) and \( y \), leveraging the open-domain nature of these coordinates, and one LT in time \( t \), which transforms the problem into spectral space \( s \) while incorporating the I.C. As demonstrated in this work, this scheme allows for straightforward computation of the Heat Kernel.  

The B.C. in the open-domain \( x \) and \( y \) directions are given by:  
\begin{equation}
T \to 0 \quad \text{as} \quad x, y \to \infty.
\end{equation}  

Cooling B.C. in the \( z \)-direction enforce NLOC at the two boundaries. See Appendix \ref{Cooling Boundary Conditions} for a detailed derivation and further discussion. Instead of using a single heat transfer coefficient \( h \), we introduce \( H_1 \) and \( H_2 \) to describe heat transfer at the left and right boundaries, respectively. For simplicity and without loss of generality, we set the ambient temperature \( T_{\infty} = 0 \), leading to:

\begin{equation}\label{NLOC}
\left. \frac{1}{T} \frac{\partial T}{\partial z} \right|_{z=0} = H_1, \quad \quad \quad
\left. \frac{1}{T} \frac{\partial T}{\partial z} \right|_{z=w} = -H_2.
\end{equation}
\vspace{1mm}

Importantly, the positive sign on the left and negative on the right enforce heat flow \textit{outward} from the sample at both ends. For further discussion on cooling B.C. and their extension to SB radiation, refer to Appendix \ref{Cooling Boundary Conditions}.

\subsection{Integral Transform Approach for the 3+1 Heat Kernel}\label{Integral Transform Approach for the 3+1 Heat Kernel}

To solve the heat equation, we apply two FT in \(x\) and \(y\) as well as one LT in $t$ on $T$, see Appendix \ref{FS and LT} for convention in use. This yields the transformed temperature profile $\hat{T}$:

\begin{widetext}
\begin{equation}\label{2+1 transformations}
T(t, \vec{x}) = \frac{1}{(2\pi)^2} \int_{-\infty}^{\infty} dk_x \, e^{i k_x \bar{x}} \int_{-\infty}^{\infty} dk_y \, e^{i k_y \bar{y}} \int_{\epsilon-i\infty}^{\epsilon+i\infty}  ds \, e^{s\bar{t}} \hat{T}(s, k_x, k_y, z),
\end{equation}

where \( \bar{x} = x - x' \), and \( \hat{T}(s,k_x,k_y,z) \) represents the temperature profile in Fourier-Laplace space. The PDE \eqref{equation in co moving frame} in the co-moving frame transforms into a second-order linear ODE in zz:

\begin{equation}\label{Eqz}
\begin{gathered}
 \gamma \hat{T} - \frac{\partial^2 \hat{T}}{\partial z^2} = \frac{1}{\alpha} \left[\hat{S}(s, k_x, k_y, z) + T(t=0,k_x,k_y,z) \right], \\
 \gamma \equiv \frac{s}{\alpha} + (k_x^2 + k_y^2) -iv k_x.
\end{gathered}
\end{equation}
The I.C. can be freely chosen to model various scenarios relevant to welding and AM processes (see Appendix \ref{IC Section} for more examples in which I.C. are important). Nevertheless, for simplicity, we set $T_{(t=0)}=0$. The solution to \eqref{Eqz} is obtained using standard Green’s function techniques, denoted by $\tilde{G}(z, z')$; see Appendix \ref{Derive GF} for details:

\begin{equation}\label{greenz}
\begin{gathered}
\tilde{G}(z, z') = \frac{1}{2 \beta (\eta \xi - 1)} \bigg[ \eta \, e^{\beta (z_> + z_<)} + e^{\beta (z_> - z_<)} + \eta \, \xi \, e^{-\beta (z_> - z_<)} + \xi \, e^{-\beta (z_> + z_<)} \bigg],\\
\eta = \frac{\beta + H_1}{\beta - H_1},\quad\quad\quad
\xi = \frac{\beta + H_2 }{\beta - H_2} \, e^{2\beta w}.
\end{gathered}
\end{equation}
where $\beta\equiv\gamma^{1/2}$, $z_{>}=\text{max}(z,z')$ and $ z_{<}=\text{min}(z,z')$. However, since both \( \eta(k_x,k_y,s) \) and \( \xi(k_x,k_y,s) \) depend on \( k_x, k_y \), and \( s \), solving the inverse Laplace transform (ILT) is highly nontrivial. In general, the integrals in  \eqref{2+1 transformations} do not commute, making direct evaluation challenging.
\subsubsection{Commuting the integrals}\label{Commuting the integrals}
To simplify the problem, we apply a useful transformation, redefining \( s \rightarrow s - \alpha(k_x^2 + k_y^2) \), which consequently shifts \( \epsilon \rightarrow \tilde{\epsilon} \) and \( \beta \rightarrow s^{1/2} \):

\begin{equation}\label{greenxyz}
\begin{aligned}
    G_{v}( \bar{x},\bar{y},\bar{t} , z, z') &= \int_{-\infty}^{\infty} dk_x \,  e^{i k_x (\bar{x}-v\bar{t})} e^{-\alpha k_x^2 \bar{t}} 
     \int_{-\infty}^{\infty} dk_y \, e^{i k_y \bar{y}} e^{-\alpha k_y^2 \bar{t}}   \int_{\tilde{\epsilon}-i\infty}^{\tilde{\epsilon}+i\infty}  ds \, e^{s \bar{t}} \tilde{G}(z,z'),\\
   &=\frac{\pi}{\alpha\bar{t}} \, e^{-\frac{[\bar{x}-v\bar{t}]^2 + \bar{y}^2}{4\alpha\bar{t}}}\int_{\tilde{\epsilon}-i\infty}^{\tilde{\epsilon}+i\infty} ds \, e^{s \bar{t}} \tilde{G}(z,z') ,\\
     &= G_{0 1D}(\bar{y}, \bar{t}) \, G_{v 1D}(\bar{x},\bar{t}) \, g(z,z',\bar{t}). \\
\end{aligned}
\end{equation}
\end{widetext}

Here, \( G_{0 1D} \) and \( G_{v 1D} \) represent the Green’s functions for a one-dimensional open domain with static and moving heat sources, respectively. Thus, our goal is to determine the effective Green’s function \( g(z,z',\bar{t}) \) in the finite-width domain.

The transformation we have applied aligns with our intuition that the dependence on \( x \) and \( y \) should factor out and correspond to the Heat Kernel in 2D. Mathematically, as demonstrated in Appendix \ref{Poles and Singular structure}, all poles of the contour integral lie on the negative real axis, \( s_{R} \).

Since the shift in \( \epsilon \) is positive definite, as given by:
\[
\tilde{\epsilon}= \epsilon + \alpha (k_x^2 + k_y^2),
\]
regardless of variations in \( k_x \) and \( k_y \), the shift in \( \epsilon \) only translates the contour integral to the right, leaving the final integral result unchanged. Consequently, we can commute the Fourier and Laplace integrals.

This physical intuition and mathematical proof will be further illustrated by comparing pure numerical schemes used to solve \eqref{equation in co moving frame} with the numerical evaluation of the integral, see Figures \ref{fig:comparison1}, \ref{fig:comparison2}, and \ref{comparison3}, as well as the alternative series approach introduced later in this work.

\subsubsection{Branch Cut due to \( s^{1/2} \)}\label{Branch Cut}

Next, the ILT is further complicated by the presence of a branch point at \( s=0 \) due to \( s^{1/2} \), introducing a branch cut that reflects the parabolic nature of the PDE, as explained earlier. Naturally, incorporating a branch cut into our contour integral requires careful handling; nevertheless, we must still evaluate the integration without issue. We choose the branch cut to be \( (-\infty,0] \). The corresponding contour is illustrated in Fig.~\ref{LT contour}.

\subsection{Evaluating \texorpdfstring{$g(z,z',\bar{t})$}{g(z,z',t-bar)}}\label{Evaluating}
Substituting \eqref{greenz} into \eqref{greenxyz}, we obtain four integrals for computing \( g(z,z',\bar{t}) \): 

\begin{widetext}
\begin{equation}\label{Isintegral}
\begin{gathered}
g(z,z',\bar{t}) = I_{(+,+)}[z_> - z_<] + I_{(+,-)}[(w - z_>) + (w - z_<)] + I_{(-,+)}[z_> + z_<] + I_{(-,-)}[(w - z_>) + (w + z_<)],\\
I_{(\pm,\pm)}(\mathcal{Z}) = \int_{\tilde{\epsilon} - i \infty}^{\tilde{\epsilon} + i \infty} \frac{1}{2 s^{1/2}} \frac{e^{s \bar{t}} e^{-s^{1/2} \mathcal{Z}} (s^{1/2} \pm H_1)(s^{1/2} \pm H_2)}{(s^{1/2} + H_1)(s^{1/2} + H_2) - (s^{1/2} - H_1)(s^{1/2} - H_2)e^{-2 s^{1/2}w} } \, ds.
\end{gathered}
\end{equation}
\end{widetext}

Here, \( \mathcal{Z} > 0 \) carries the dependence on \( z_< \) and \( z_> \), while \( \alpha \) has been absorbed \( s/\alpha \rightarrow \alpha\) for conciseness and will recovered later, without loss of generality. We also suppress the explicit \( t \)-dependence of \( I_{(\pm,\pm)} \) for conciseness. 

The key observation is that the four integrals \( I_{(\pm,\pm)} \) share the same singular structure, which is independent of \( z \) and depends only on the cooling coefficients \( H_1, H_2 \), thermal diffusivity \( \alpha \), and the width of the sample \( w \). Thus, it requires only a \textit{one-time computation}. A similar behavior is expected in the SB or polynomial cooling cases.

\subsubsection{Poles' Rank?}\label{Poles' Rank?}

In addition to the branch cut, this integral presents another challenge: the singular structure introduced by the term \( e^{-2 s^{1/2} w} \), which is difficult to handle analytically. Fortunately, this term does not alter the rank of the singularities—all remain simple poles, which can be easily verified numerically. 

In the absence of this term, the equation reduces to a quadratic in \( s^{1/2} \), yielding at most two simple poles. However, while \( e^{-2 s^{1/2} w} \) does not change the rank of the poles, it renders them infinitely countable with monotonically increasing spacing between them. As demonstrated in Appendix \ref{Poles and Singular structure}, all these poles lie on the negative \( s_{R} \)-axis. We determine their positions numerically.

\subsubsection{Using the Residue Theorem}\label{Using the Residue Theorem}

\begin{figure}
\includegraphics[width=8cm]{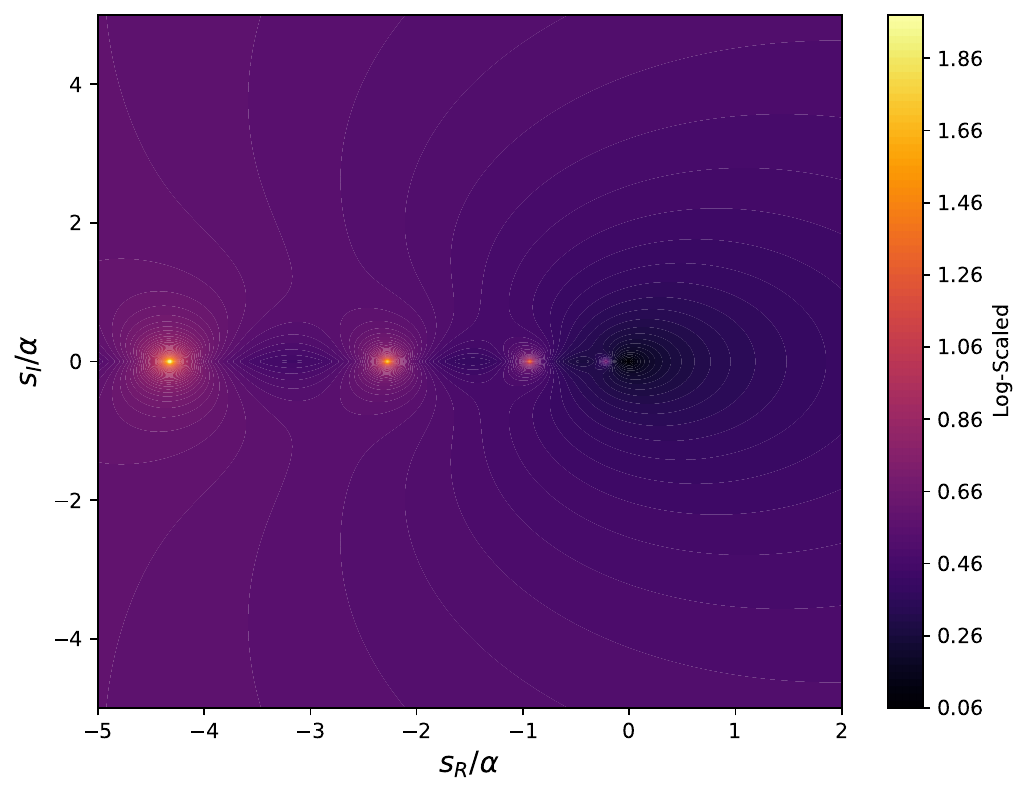}
        \caption{
           Contour plot of integrand of \eqref{Isintegral} in the complex \( s \)-plane, where \( s/\alpha = s_{R}/\alpha + i s_{I}/\alpha \). The plot highlights the simple poles for the specific choice of parameters as stated in the text. All poles lie on the negative $s_R$-axis, and not evenly spaced as proved in Appendix \ref{Poles and Singular structure}.
        }
\label{poles}
\end{figure}

Relying on the Estimation Lemma and the causality conditions on \( \mathcal{Z} \), as explained in the previous subsection, we can evaluate the contour integral using the Residue Theorem \cite{schiff1999laplace,churchill2009complex}. % minimal refs, text unchanged
\begin{widetext}
\begin{equation}\label{Soltuion eq FS}
\begin{gathered}
g(z,z',\bar{t})=\sum_{n=1} A(\beta_n) e^{-\beta^2_n \bar{t}} \left\{ H_2 \sin\left[ \bar{z}_{>}  \sqrt{\beta_n} \right] + \sqrt{\beta_n} \cos\left[ \bar{z}_{>}  \sqrt{\beta_n} \right] \right\}\left\{ H_1 \sin\left[ z_{<} \sqrt{\beta_n} \right] + \sqrt{\beta_n} \cos\left[ z_{<} \sqrt{\beta_n} \right] \right\}\\
A(\beta_n) = \frac{2 \alpha \sqrt{\beta_n}}{\sin\left[ w \sqrt{\beta_n} \right] \left\{ \beta_n \left[ (H_1 + H_2) w + 3 \right] - H_1 H_2 \right\} + \sqrt{\beta_n} \cos\left[ w \sqrt{\beta_n} \right] \left\{ w \left[ \beta_n - H_1 H_2 \right] - 2 (H_1 + H_2) \right\}}\\
\end{gathered}
\end{equation}
where \( \bar{z} \equiv w - z \), $\beta_n=\sqrt{s_n/\alpha}$ and we have reinstated the \( \alpha \) factor for completeness. In \eqref{Soltuion eq FS}, we performed the transformation \( s_n \rightarrow -s_n \), ensuring that \( s_n \) remains positive definite. This avoids ambiguity from taking square roots in spectral space and makes the exponential damping explicit. We list some of the poles for the chosen parameters $w=5\, mm$, $H_1=1.2 \, W.mm^2/K$ and $H_2=1 \, W.mm^2/K$ in Table~\ref{Poles}, with their distribution illustrated in Fig.~\ref{poles}.

\begin{figure}[H]
    \centering
    \begin{minipage}{0.48\textwidth}
        \centering
        \includegraphics[width=\textwidth]{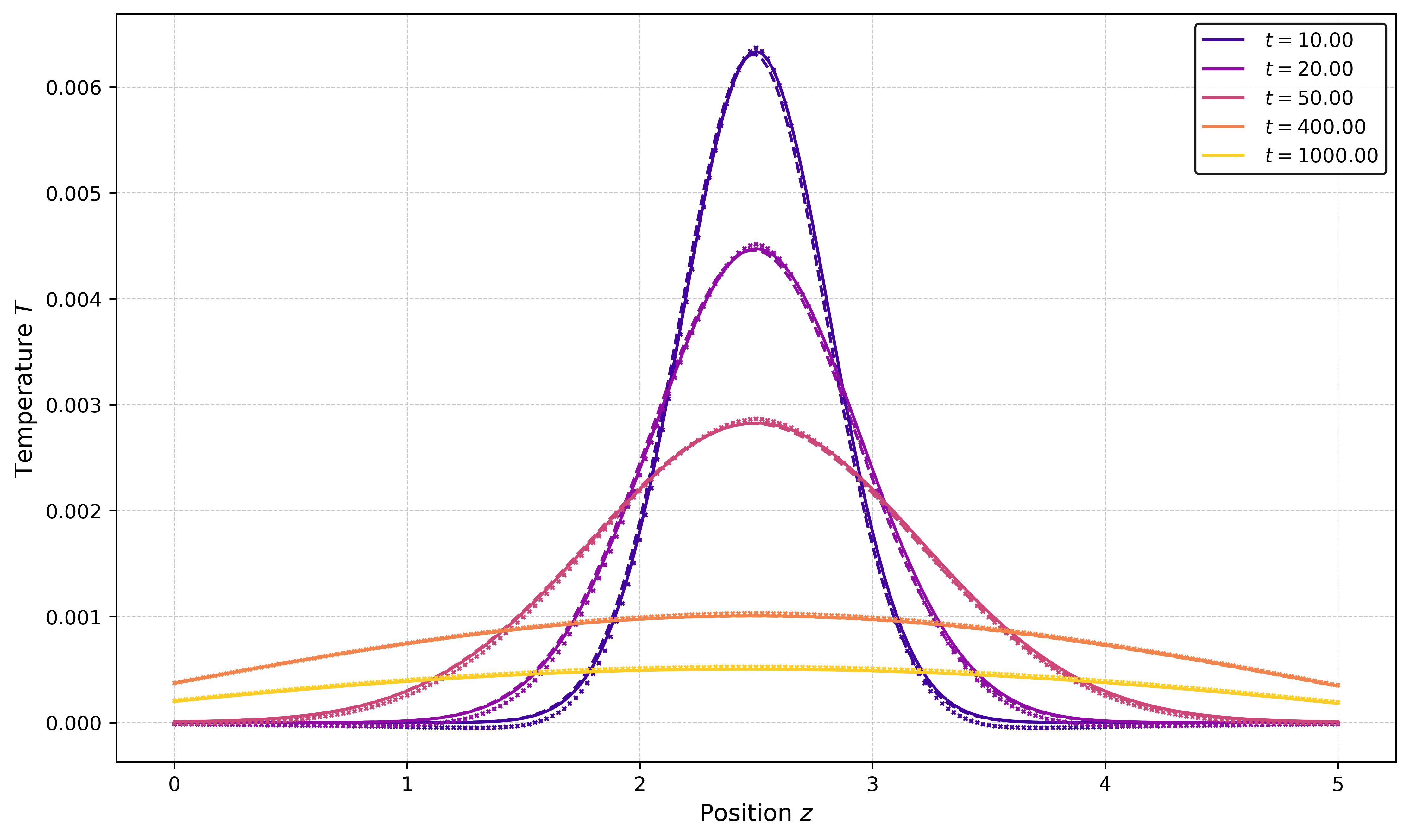}
        \caption{
            Comparative analysis of temperature profile evolution using three distinct solution methods over time. Dashed Line: Numerical solution obtained by inverting the LT with the \textsc{MP-python} package. Solid Line: Solution generated by the \textsc{EWI-NumericalHeatSolver} package using the Finite-Difference (FD) method on a uniform mesh. Crossed Line: Analytical series solution, truncated after 160 terms, reported in this article. The alignment of these solutions across different time intervals illustrates strong agreement, validating the robustness of each method in capturing temperature dynamics and accurately modeling heat propagation behavior .
        }
        \label{fig:comparison1}
    \end{minipage}
    \hfill
    \begin{minipage}{0.48\textwidth}
        \centering
        \includegraphics[width=\textwidth]{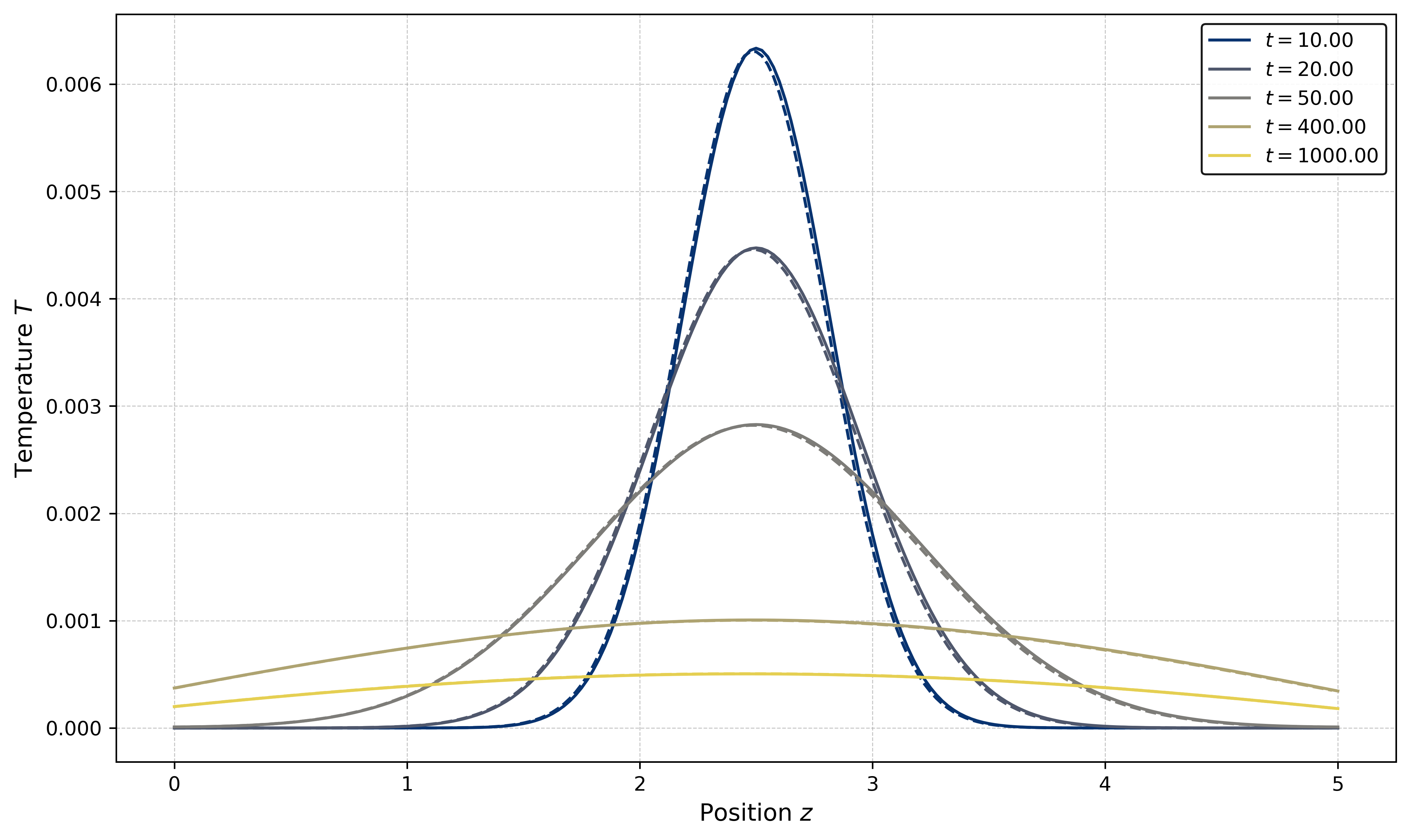}
        \caption{
            A comparison of temperature profile evolution obtained through ILT and FD methods. The dashed line represents the temperature profile derived by numerically inverting the LT using the \textsc{MP-python} package. The solid line shows the solution generated using the \textsc{EWI-NumericalHeatSolver} package, implemented via the FD method on a uniform mesh. This comparison demonstrates a strong agreement between the two approaches across the entire temperature profile evolution, validating the analytical model implemented in this work.
        }
        \label{fig:comparison2}
    \end{minipage}
\end{figure}

\end{widetext}

\begin{table}[htbp]
\centering
\begin{tabular}{|c|c|}
\hline
$n$ & $-s_n / \alpha$ \\
\hline
1  & 0.21788344122246012 \\
2  & 0.9354984090489751 \\
3  & 2.274537277073521 \\
4  & 4.323471861197404 \\
5  & 7.125975596988451 \\
6  & 10.701046303359455 \\
7  & 15.057037950335976 \\
8  & 20.197859113993204 \\
9  & 26.125475142527932 \\
10 & 32.840943281468654 \\
11 & 40.344866943627416 \\
12 & 48.63760847150312 \\
13 & 57.71939510962397 \\
\vdots & \vdots \\
\hline
\end{tabular}
\caption{For the NLOC cooling B.C, this table show the locations of first few poles in spectral space \( s \), normalized to \( \alpha \),  for a specific choice of parameters as stated in the text.}
\label{Poles}
\end{table}

\subsubsection{Causality Structure}\label{Causality Structure}
Moreover, the temperature profile \( T \), obtained through the ILT, exhibits a causality structure that reflects the retardation of the Green’s function. This causality condition ensures that the solution is valid only after the impact, i.e., for \( t > t' \), through the step function \( \Theta(t - t') \). Nevertheless, for conciseness, we do not explicitly write down the step functions in the previous expressions \eqref{Isintegral}.

\subsection{Alternative Approach: Fourier Series}\label{Alternative Approach: Fourier Series}
\begin{figure}
\includegraphics[width=8cm]{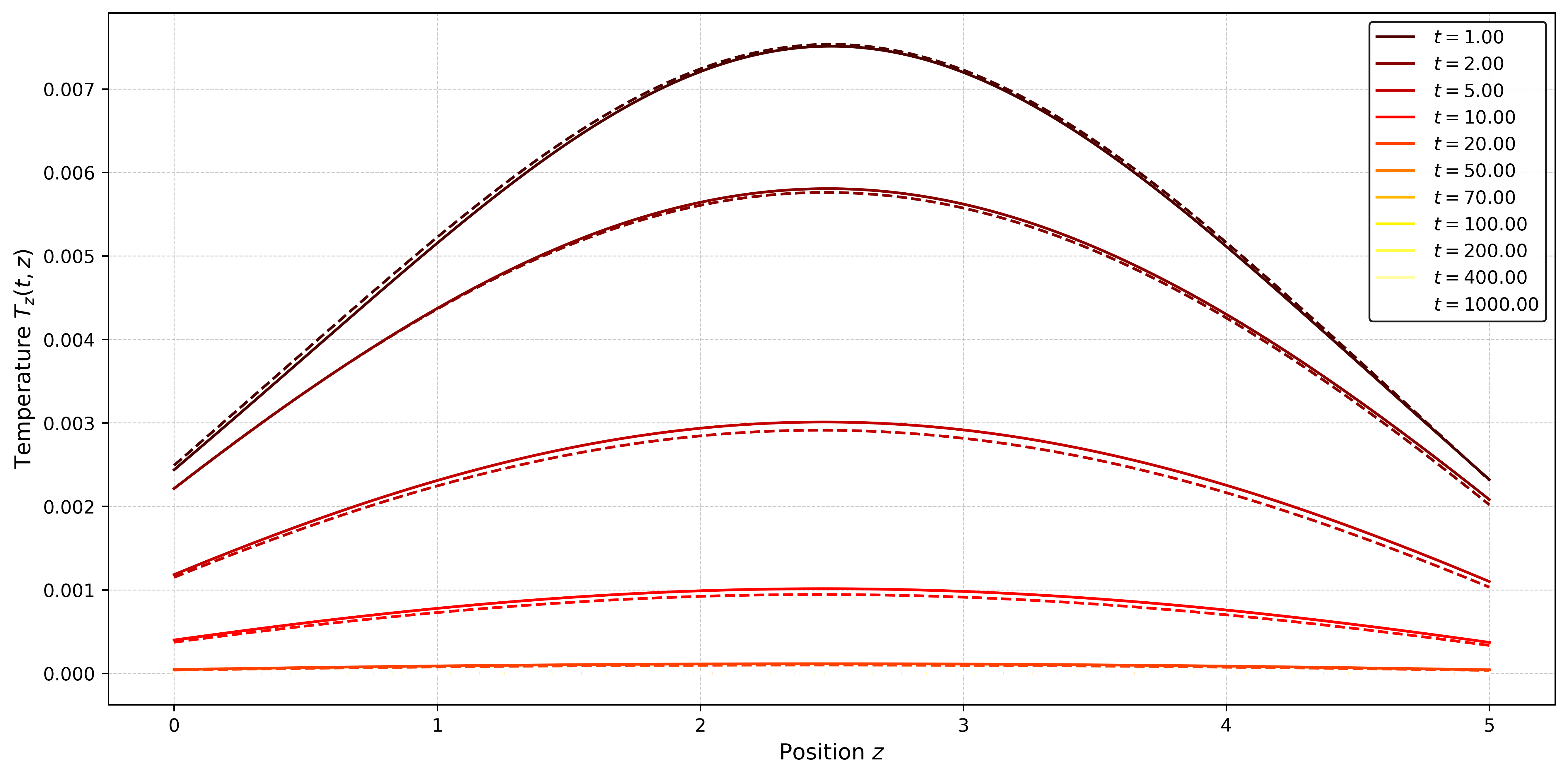}
\caption{A comparison of temperature profile evolution obtained through ILT and FD method. The dashed line represents the temperature profile derived by numerically inverting the Laplace integral using the \textsc{MP-python} package. The solid line shows the solution generated using the \textsc{EWI-NumericalHeatSolver} package, implemented via the FD method on a uniform mesh. This comparison demonstrates a strong agreement between the two approaches across the entire temperature profile evolution, validating of analytical model implemented in this work.
}
\label{comparison3}
\end{figure}

In this section, we demonstrate that there is an alternative approach to obtain the Green function \( G \), involving the use of the FS instead of the LT, see Figures ~\ref{fig:comparison1} and ~\ref{comparison3} for a visual comparison between solutions obtained using both approaches. While LT leads to a spectral representation in \( s \) and is more naturally suited for dealing with IBVPs or IVPs, FS reformulates the problem in terms of discrete number of eigenfunctions \( \phi_n(z) \), naturally accommodating finite-width constraints. It also recovers the infinite-width case in the limit of \( w \to \infty \), ultimately becoming equivalent to a the third FT used in the 3D Rosenthal model.

Although we do not provide a full proof here, we emphasize that as \( w \) increases, the spacing between poles \( \Delta s_n \equiv s_{n+1} - s_n \) should decrease. In the limit \( w \to \infty \), we obtain \( \Delta s_n \to 0 \), transforming the FS summation into an FT integral.

By expanding both the temperature \( T \) and source term \( S \) in terms of those eigenfunctions \( \phi_n(z) \), the heat equation reduces to an effective first-order ODE in \( t \), analogous to the LT approach. The final Green’s function representation in this formulation takes the form:

\begin{equation}
\begin{aligned}
G(t, x, y, z, x', y', z') &= G_{0,1D}(\bar{y}, \bar{t}) G_{v,1D}(\bar{x}, \bar{t}) \Theta(\bar{t})\\
&\quad \times \sum_n \phi_n(z) \phi_n(z') e^{- \lambda_n^2 \bar{t}}.
\end{aligned}
\end{equation}

A detailed derivation of this result, including the eigenvalue equation and B.C, is provided in Appendix \ref{FS Derivation}. To illustrate the equivalence, in Figures ~\ref{fig:comparison1}, ~\ref{fig:comparison2}, ~\ref{comparison3}, and ~\ref{fig:10plots}, we compare among the pure numerical implementation with our analytical solution obtained using both the FS and LT approaches.

\subsection{Brief Summary}\label{Brief Summary}
In this section, we addressed heat transfer in a semi-infinite material with cooling governed by NLOC B.C. We obtained the Green function \( G_v \) as a closed-form analytical solution, incorporating finite-geometry effects. 

\begin{figure}
\includegraphics[width=8cm]{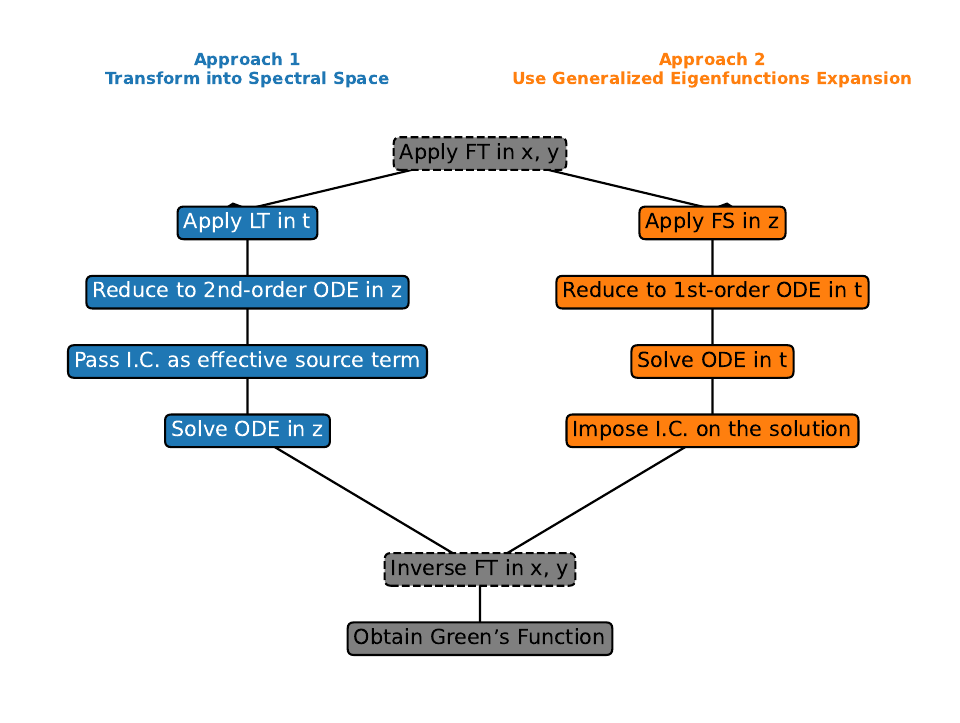}
        \caption{
           This figure compares two approaches for obtaining the Green’s function. Both methods start with FT in \( x \) and \( y \). The first approach applies a LT in \( t \), reducing the problem to a second-order ODE in \( z \), while the second approach uses a FS in \( z \), leading to a first-order ODE in \( t \), which is solved before imposing I.C. In both methods, the Inverse FT in \( x \) and \( y \) independently of analysis needed for solving the effective ODEs. The Green's function obtained using both methods are equivalent.}
\label{SketchFor Apporaches}
\end{figure}

To solve this system, we introduce two equivalent approaches, as sketched in Fig.~\ref{SketchFor Apporaches}. The first method applies two FT in \( x \) and \( y \) and an LT in \( t \), reducing the \( 3+1 \) heat equation to a second-order ODE in \( z \), which we solve using Green’s function techniques in one dimension. However, inverting the LT is nontrivial due to poles and a branch cut, requiring careful handling of the contour integration. The second approach replaces the LT with an FS expansion in \( z \), reformulating the problem in terms of discrete eigenfunctions. This leads to a first-order ODE in \( t \), which is solved using direct integration before imposing I.C, recovering the continuous FT in the limit \( w \to \infty \).

The key objective is to use this analytical Green’s function as the foundation for constructing heat profiles for realistic sources while incorporating cooling and finite boundary constraints.

\section{Heat Profile for different Source terms}\label{Heat Profile for different Source terms}
At this point, we are well-prepared to study the heat profile for more interesting source terms in the co-moving frame. With the kernel at hand, we can solve for the heat profile by integrating with the source term:
\begin{widetext}

\begin{equation}\label{Convlution}
\begin{aligned}
T(t, x, y, z) &=  \int_{-\infty}^{t} dt' \int_{0}^{w} dz' \int_{-\infty}^{\infty} dx' \int_{-\infty}^{\infty} dy' \, G_{v}(\bar{x}, \bar{y}, \bar{t}, z, z') S(t', x', y', z').
\end{aligned}
\end{equation}

\begin{figure}[ht]
    \centering
    % First row
    \begin{minipage}{0.3\textwidth}
        \centering
        \includegraphics[width=\textwidth]{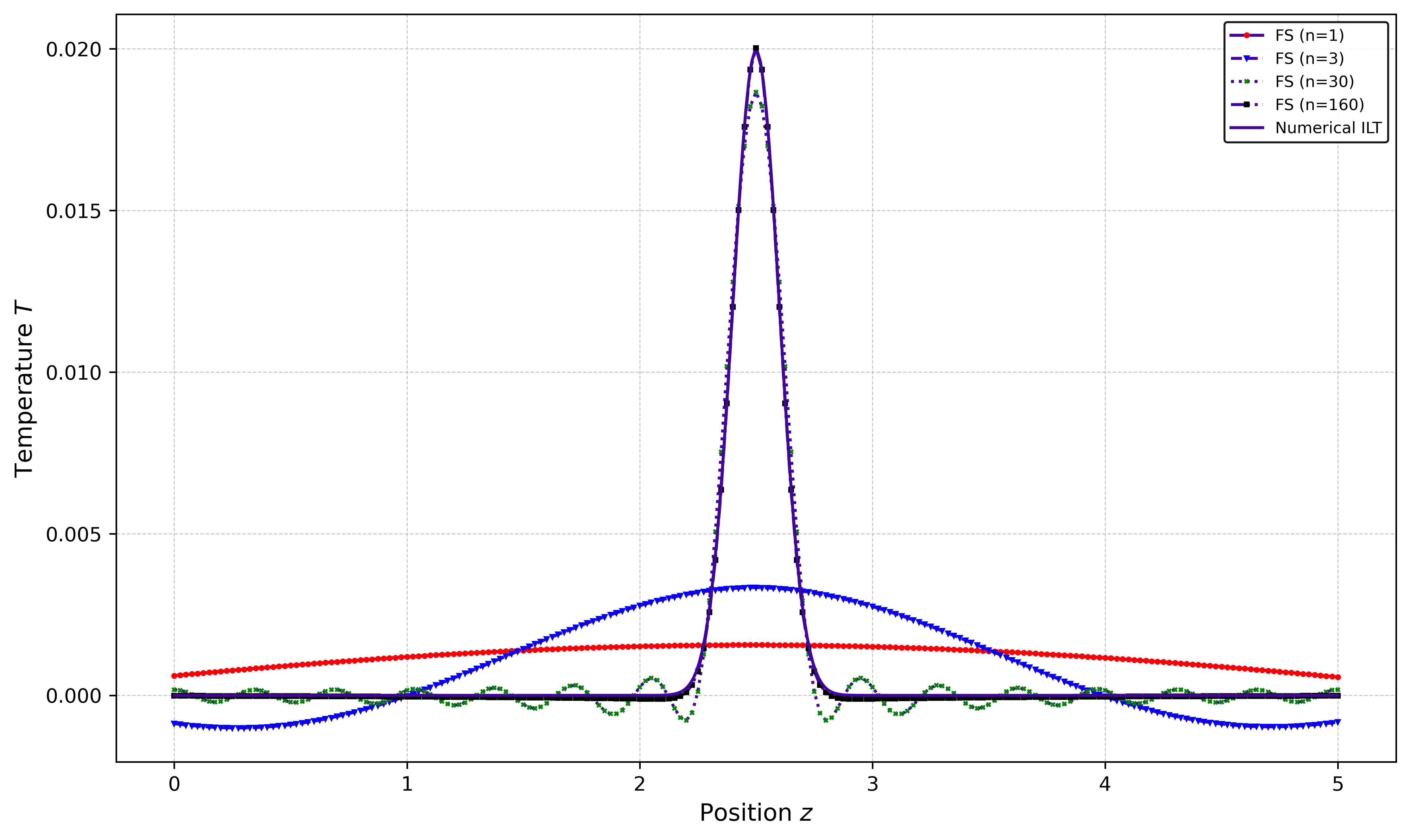}
        \subcaption{$t = 1$}
    \end{minipage}
    \hfill
    \begin{minipage}{0.3\textwidth}
        \centering
        \includegraphics[width=\textwidth]{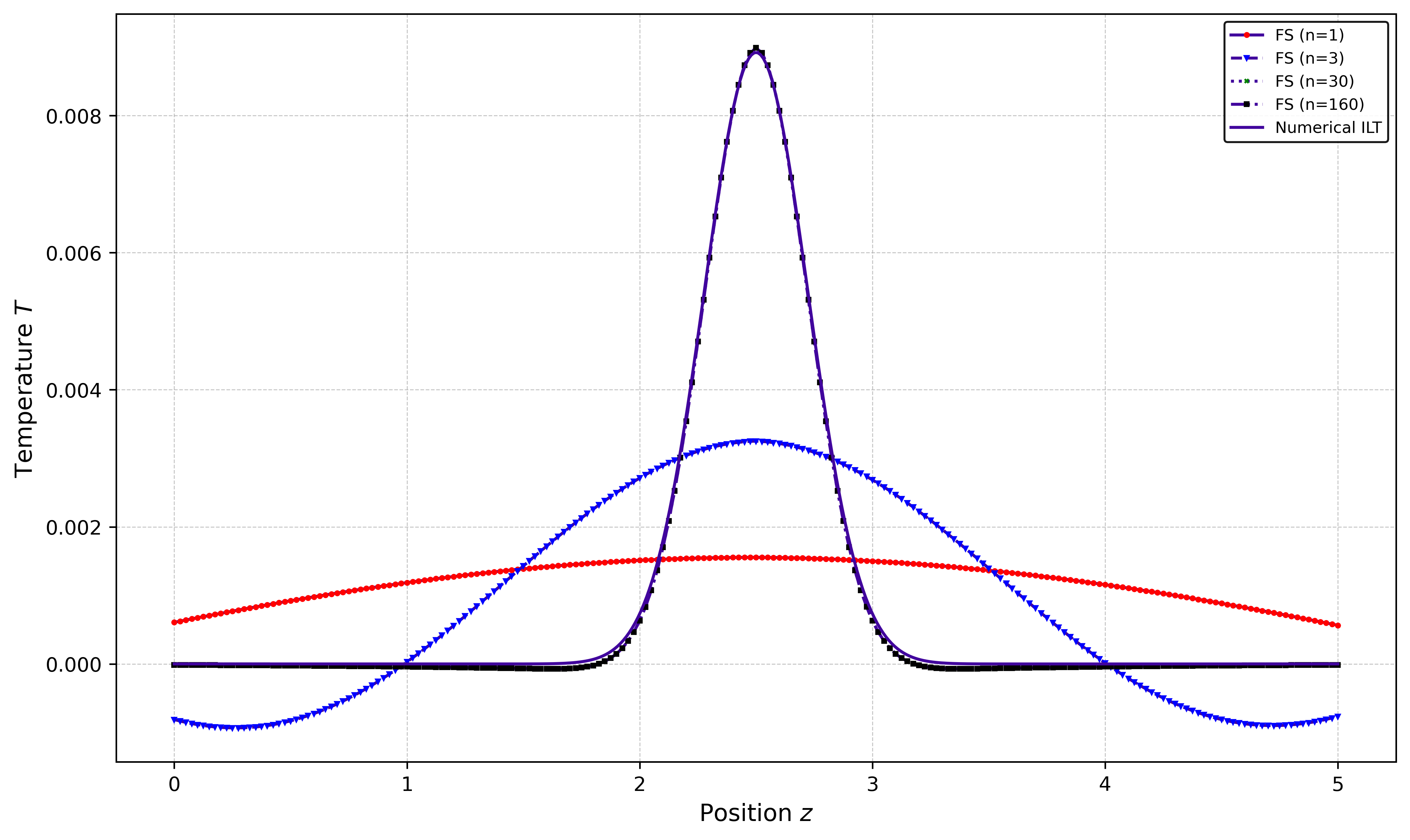}
        \subcaption{$t =5$}
    \end{minipage}
    \hfill
    \begin{minipage}{0.3\textwidth}
        \centering
        \includegraphics[width=\textwidth]{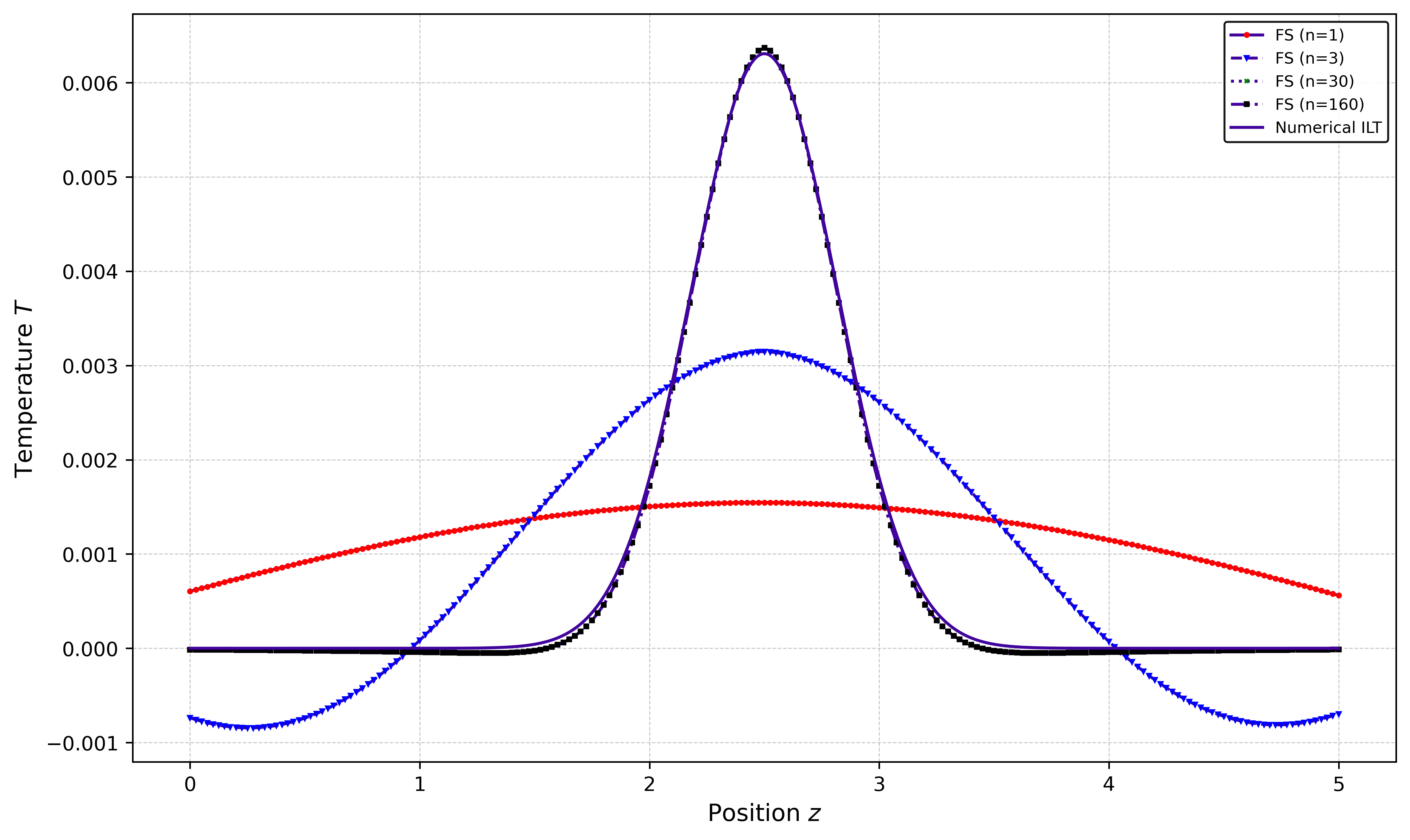}
        \subcaption{$t = 10$}
    \end{minipage}
    
    \vspace{0.4cm} % Space between rows
    
    % Second row
    \begin{minipage}{0.3\textwidth}
        \centering
        \includegraphics[width=\textwidth]{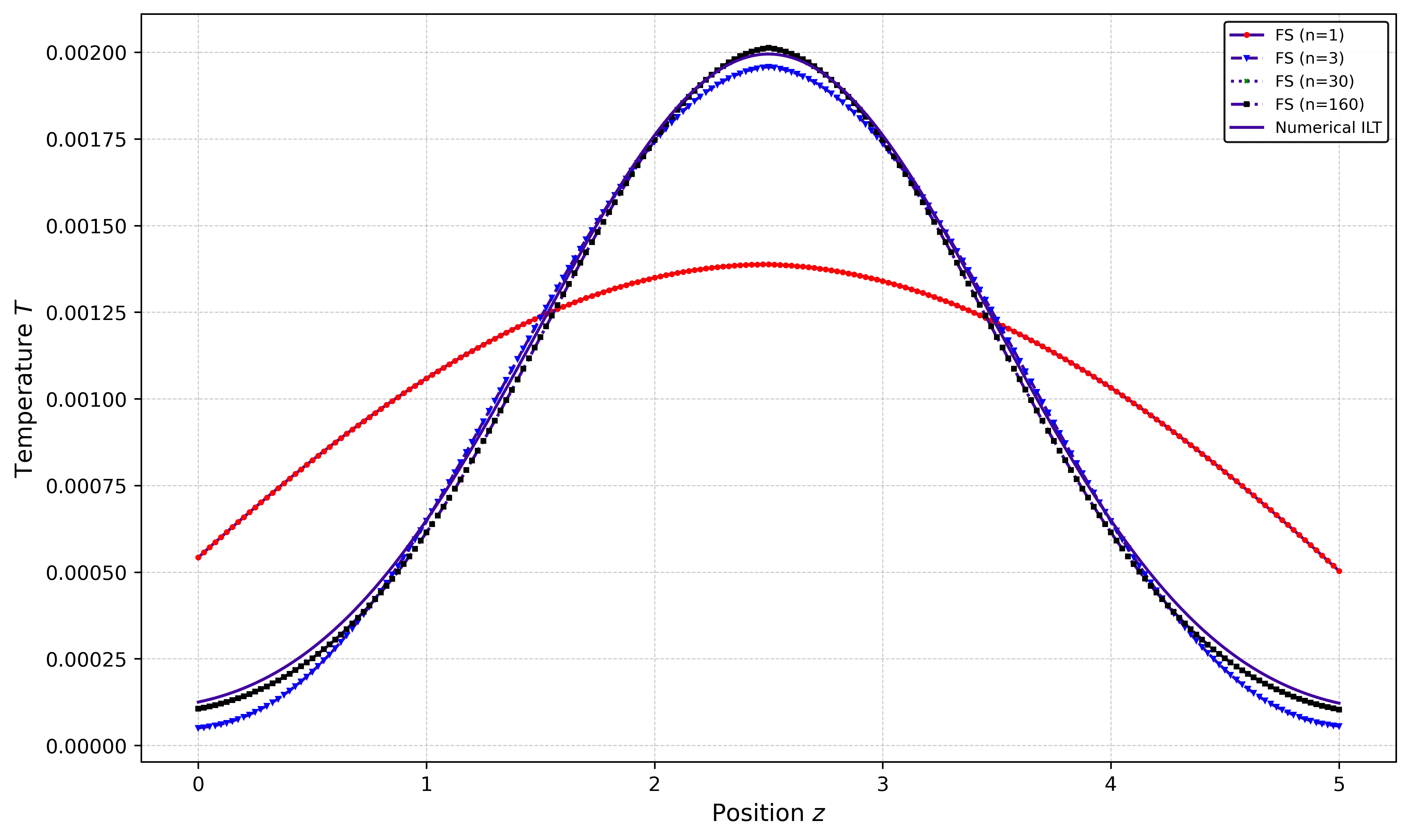}
        \subcaption{$t = 100$}
    \end{minipage}
    \hfill
    \begin{minipage}{0.3\textwidth}
        \centering
        \includegraphics[width=\textwidth]{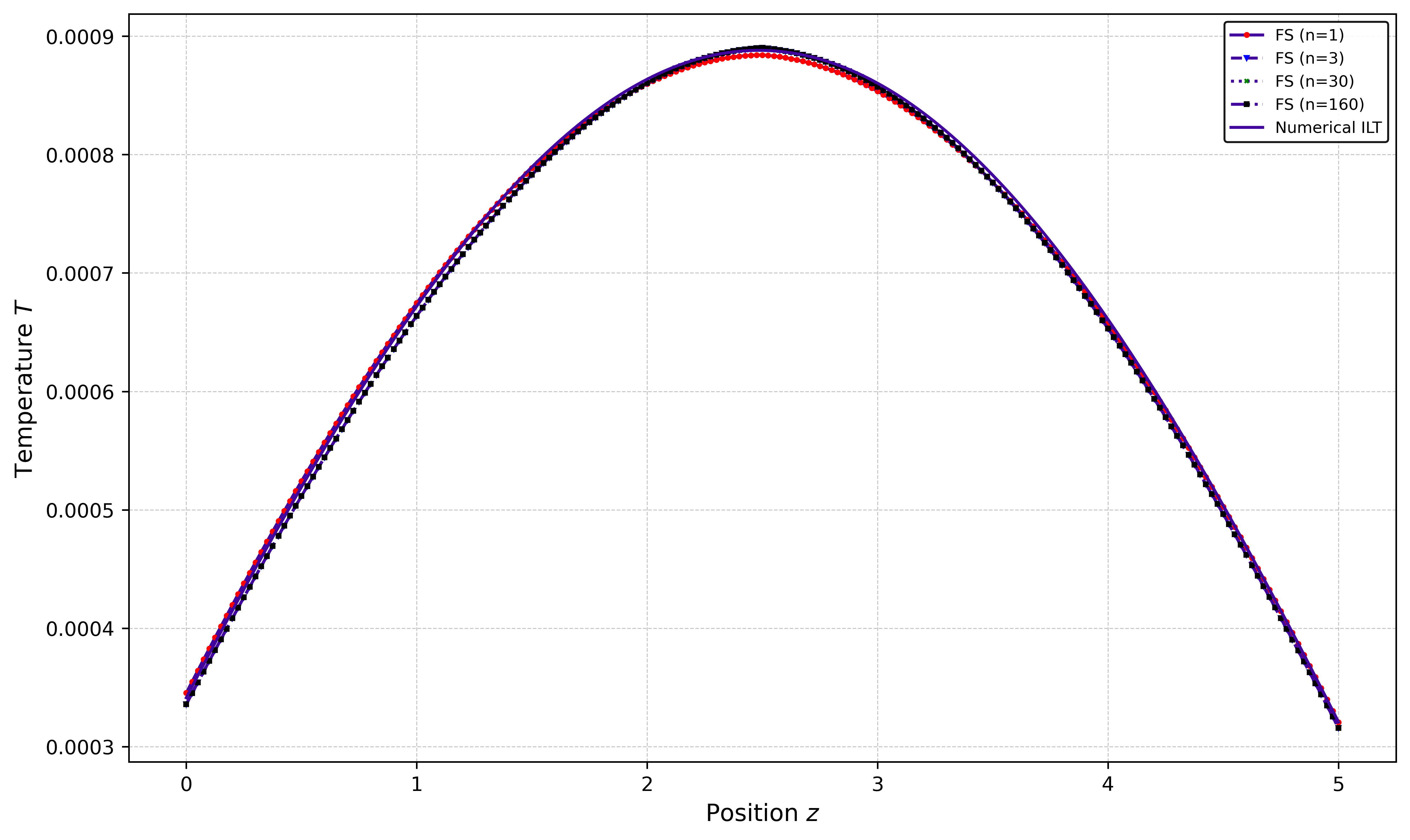}
        \subcaption{$t = 500$}
    \end{minipage}
    \hfill
    \begin{minipage}{0.3\textwidth}
        \centering
        \includegraphics[width=\textwidth]{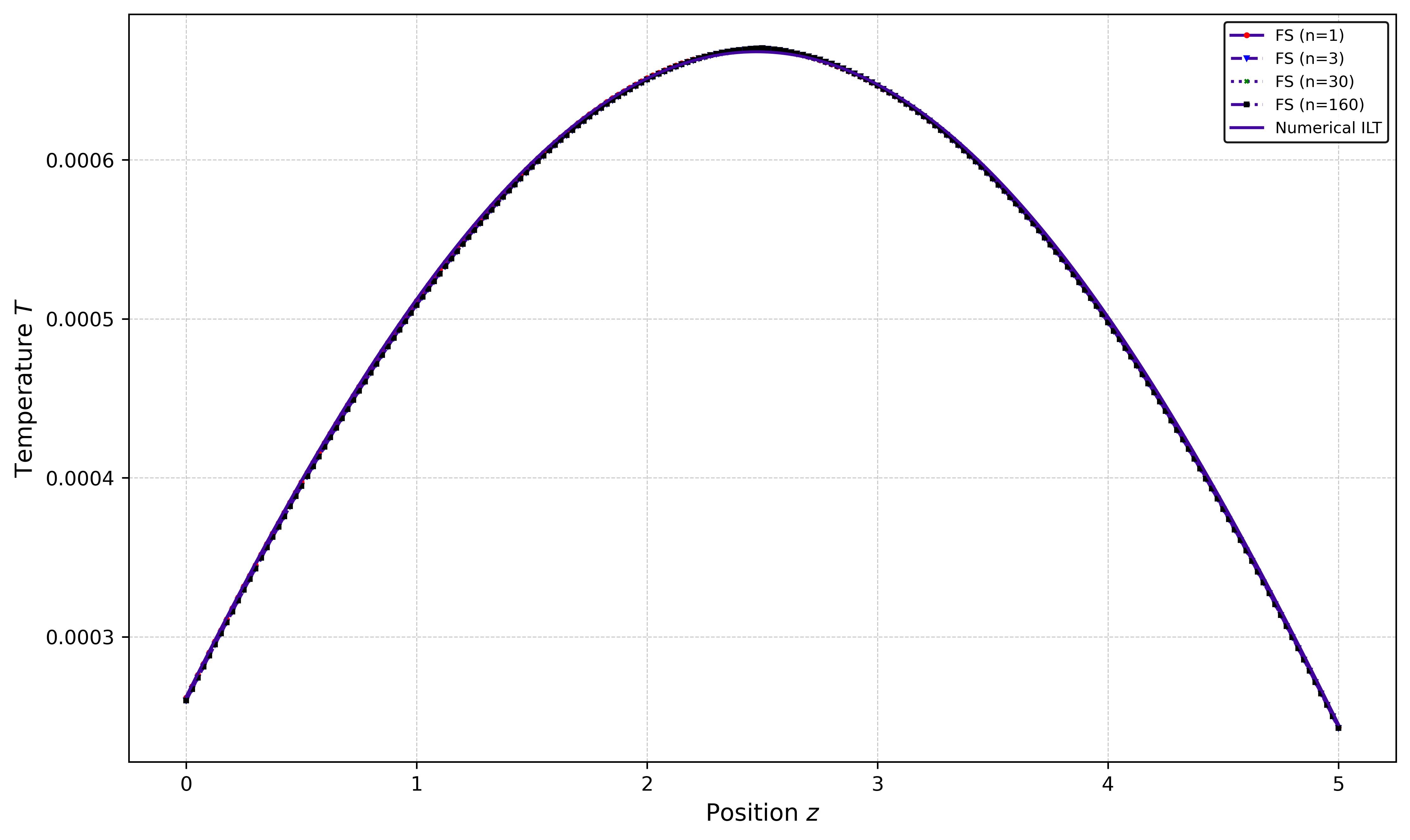}
        \subcaption{$t = 750$}
    \end{minipage}
    
    \caption{This figure compares temperature profiles derived from the implementation of the ILT and truncated FS solutions reported in this article. The series solutions are truncated after $n$-terms: $ 1, 3, 30$, and $160$ and evaluated at various time intervals: \( t = 1, 5, 10, 100, 500,\) and \( 750 \). The comparison shows that the series solutions converge over time due to the rapid decay of higher-order modes compared to the fundamental mode. At later times, the profile becomes increasingly dominated by the fundamental mode as cooling effects fully manifest. This suggests that for late-time evolution, only a few terms are needed to accurately represent the profile, as the influence of higher modes diminishes significantly after cooling is initiated.}
    \label{fig:10plots}
\end{figure}
\end{widetext}

We have two options when it comes to evaluate this integral using $G_{v}$:

\begin{enumerate}
    \item \textbf{Commuting the Integrals:} Interchanging the order of ``source" integral with spectral one. If the former one is easier to execute under the ILT, potentially leading to closed-form solutions or reducing the problem to a more manageable one-dimensional integral. On the other hand, this approach might change the pole structure of the spectral integral.

    \item \textbf{Integration inside the Series:} Whether we use LT or FS, we end up with a series solution. Thus, we can integrate each term individually offers a more straightforward computational approach compared to the first method.
\end{enumerate}

Each method has trade-offs, and the best choice depends on the source’s form. In the following subsections, we will focus on the later option. For examples of the former one, please refer to the Appendix \ref{source Using LT} and \ref{Using LT OPTION1}. 

Let's assume a factorized form of the source term \( S(t, x, y, z) = S_{x}(t, x) S_{y}(t, y) S_{z}(t, z) \), in the co-moving frame, then we can re-write the integral as follows:
\begin{widetext}
\begin{equation}\label{FactroizedSourceConv}
\begin{aligned}
T(t, x,y,z) &=  \int_{-\infty}^{t}dt' \, \int_{0}^{w} dz' g(z,z', \bar{t}) S_{z} \, \int_{-\infty}^{\infty} dx' G_{v 1D}(\bar{x}, \bar{t}) S_{x} \, \int_{-\infty}^{\infty} dy' G_{0 1D}(\bar{y}, \bar{t})  S_{y}\\
&= \int_{-\infty}^{t}dt' \,T_x(\bar{t},x)T_y(\bar{t},y)T_z(\bar{t},z)
\end{aligned}
\end{equation}
\end{widetext}
At this stage, we can proceed by exploring specific cases of the source terms and attempt to derive the corresponding temperature profile \( T \).

\subsection{Gaussian and Ellipsoidal Profiles}\label{Gaussian and Ellipsoidal Profiles}
Consider a heat source that deposits a constant power density \( A_0 \) with a spatial distribution given by:
\begin{equation}
S(x,y,z) = A_0 e^{\left(-\frac{(x-x_c)^2}{2\sigma_x^2}-\frac{(y-y_c)^2}{2\sigma_y^2}-\frac{(z - z_c)^2}{2\sigma_z^2}\right)}.
\end{equation}
In general, \( A_0 \) could be \( t \)-dependent, but we assume it to be constant for simplicity. Extending to \( A_0 \to A(t) \) is straightforward, as discussed later in \ref{Comment on Time-Dependent Source}. Without loss of generality, we set \( x_c = y_c = 0 \). A Gaussian distribution is recovered in the limit:
\[
\sigma_x = \sigma_y = \sigma_z.
\]

Integrating over \( x \) and \( y \) follows directly from Rosenthal’s result\cite{rosenthal1946} :
\begin{widetext}
%typo here,the exponet denominator might be not correct 
\begin{equation}\label{Gaussian in x}
\begin{gathered}
T_{x}(t,x)=\frac{2 \sqrt{\pi A_0} \sigma_x  \exp \left( -\frac{\left[ x - v \bar{t} \right]^2}{2 \left[ 2 \bar{t} \alpha + \sigma_y^2 \right]} \right)}{\sqrt{2 + \frac{\sigma_x^2}{\bar{t} \alpha}}} \quad \quad \quad \quad T_{y}(t,y)=\frac{2 \sqrt{\pi A_0}  \sigma_y \exp \left( -\frac{y^2 }{2 \left[ 2 \bar{t} \alpha + \sigma_y^2 \right]} \right)}{\sqrt{2 + \frac{\sigma_y^2}{\bar{t} \alpha}}}
\end{gathered}
\end{equation}
On the other hand, the integration in \( z \) is more involved and takes the following form:

\begin{equation}\label{ZPartSeriesSolution}
\begin{gathered}
    T_z(t,z) =\frac{\sigma_z }{2 \sqrt{2}} \sum_{n} e^{-\frac{s_n \bar{t}}{\alpha}-\frac{s_n \sigma_z^2}{2\alpha}} \left( \mathcal{B}_{n}(z) \left\{ \beta_n \cos\left[\bar{z} \beta_n\right] + H_2 \sin\left[\bar{z} \beta_n\right] \right\} 
    - \mathcal{C}_{n}(z) \left\{ \beta_n \cos(z \beta_n) + H_1 \sin(z \beta_n) \right\} \right),\\
\end{gathered}
\end{equation}
\end{widetext}
where \( \mathcal{B}_{n}(z) \) and \( \mathcal{C}_{n}(z) \) are defined in Appendix \ref{Formulas}.																				   
\subsection{Double Ellipsoidal spatial Profiles}\label{Double Ellipsoidal spatial Profiles}
In a more realistic setup, the heat source in the co-moving frame is not fully axially symmetric, as it does not behave like a rigid body. A slightly more interesting case arises when the heat profile follows a double ellipsoidal distribution, mimicking the asymmetry between the front and back ends along the motion direction \cite{goldak1984}. % added cite; sentence unchanged

\begin{equation}
S_{x}(x) = \left[ A_R e^{-\frac{x^2}{2\sigma_{y}^2}} \Theta(x) + A_L e^{-\frac{x^2}{2\sigma_{y}^2}} \Theta(-x) \right].
\end{equation}

The integration in \( x \) can still be carried out analytically using the error function \( \text{Erf} \), as follows:

%typo here,check this expression

\begin{widetext}
\begin{equation}
\begin{aligned}
T_{x}(t,x)&=\frac{  \sqrt{\pi A_R} \sigma_{xR}}{\sqrt{2 + \frac{\sigma_{xR}^2}{ \bar{t} \alpha}}} 
\exp \left( -\frac{\left[ x -  v \bar{t} \right]^2}{2 \left[ 2 \bar{t} \alpha + \sigma_{xR}^2 \right]} \right) 
\left( 1 + \operatorname{Erf} \left\{ \frac{\left( \frac{x}{\bar{t}} - v \right)}{\sqrt{2 + \frac{\sigma_{xR}^2}{\bar{t} \alpha}}}  \frac{\sigma_{xR}}{2 \alpha} \right\} \right)\\
&+\frac{  \sqrt{\pi A_L} \sigma_{xL}}{\sqrt{2 + \frac{\sigma_{xL}^2}{\bar{t} \alpha}}} 
\exp \left( -\frac{\left[ x - v \bar{t} \right]^2}{2 \left[ 2 \bar{t} \alpha + \sigma_{xL}^2 \right]} \right) 
\left( 1 - \operatorname{Erf} \left\{ \frac{\left( \frac{x}{\bar{t}} - v \right)}{\sqrt{2 + \frac{\sigma_{xL}^2}{\bar{t} \alpha}}} \frac{\sigma_{xL}}{2 \alpha} \right\} \right)\\
\end{aligned}    
\end{equation}
\end{widetext}
We can easily recover the Gaussian profile by taking the limit \( \sigma_{xL} = \sigma_{xR} \equiv \sigma_{x} \) and \( A_R = A_L \equiv A_0/2 \). The reader can verify that \( T_x(t,x) \) then reduces to the form given in \eqref{Gaussian in x}.

\subsection{Comment on Time-Dependent Source}\label{Comment on Time-Dependent Source}
For a heat pulse, the source can be modeled as a Dirac delta function in time, eliminating the need for a one-dimensional integral over \( t \). In this case, the closed-form analytical Green’s function for the \( 3+1 \) model provides a complete solution. 

However, for a long-exposure scenario, the integration in \eqref{FactroizedSourceConv} over \( t \) becomes nontrivial. If the power remains constant over time (\( A(t) = A_0 \)), the integral retains the same structural form but simplifies quantitatively. Conversely, if the power is explicitly time-dependent (\( A(t) \neq A_0 \)), the integral undergoes a qualitative change due to the evolving source term. From a numerical perspective, both cases follow the same computational approach; the only modification is the functional form of \(A(t)\).

We urge the reader to check some examples on time dependent source in Appendix \ref{source Using LT}.
\begin{figure}[ht]
\includegraphics[width=8cm]{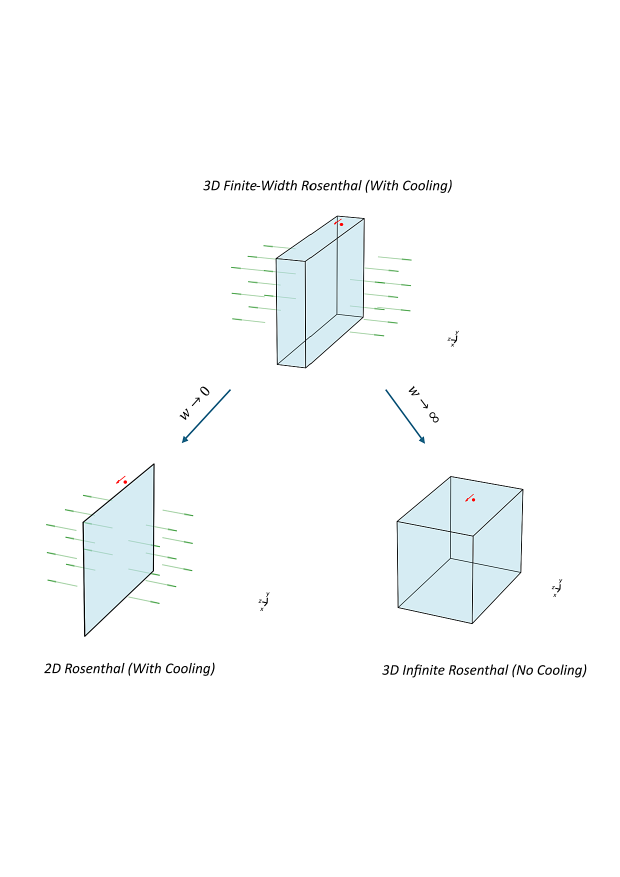}
  \caption{This figure illustrates that our cooling model, as reported in this paper, is a generalization of Rosenthal’s models in two and three dimensions. In the limit of \( w \to 0 \), our model reduces to the 2D Rosenthal model with cooling, while in the limit \( w \to \infty \), it converges to the 3D infinite Rosenthal model without cooling. Unlike these limiting cases, our model can capture variations in the \( z \)-direction and solve the problem in full three dimensions without additional assumptions. Moreover, it can be further generalized to account for finite width in both the \( x \) and \( y \) directions.}
\label{comparisonWithRosenthal}
\end{figure}

\section{Rosenthal’s Model as a Special Case of Our Framework} \label{Rosenthal_as_special_case}

In \cite{rosenthal1946}, Rosenthal developed analytical models for 2D heat transfer with cooling effects and 3D heat transfer without cooling, both under a steady-state constraint. Our model extends Rosenthal’s work in two key aspects. First, it accommodates finite-geometry 3D heat transfer while incorporating cooling effects, enabling a more comprehensive analysis in the finite-width regime. Second, it accounts for transient behavior which is crucial in understanding crack formation. In the following subsections, we compare Rosenthal’s results with our model and demonstrate that they emerge as limiting cases, as illustrated in Fig.~\ref{comparisonWithRosenthal}.

\subsection{Rosenthal's 2D Model as a Special Case at Limit of $w\rightarrow0$}\label{Rosenthal 2D}

Rosenthal's 2D cooling model is formulated under the thin-width assumption \( w \to 0 \), which corresponds to a specific limit of our more general framework, able to probe the temperature profile more in finite width $w$. To demonstrate this connection, we will work with the Temperature conformal function $\phi$ defined in \cite{rosenthal1946} as follows:
\begin{equation}
T(t,\vec{x}) = e^{-\frac{v x}{2\alpha}} \phi(t,\vec{x}).
\end{equation}

By approximating the second $z$-derivative in the corresponding Heat PDE on $\phi$:

\begin{equation}
\frac{\partial^2 \phi}{\partial z^2} \approx \frac{\frac{\partial \phi}{\partial z}\big|_{z=w} - \frac{\partial \phi}{\partial z}\big|_{z=0}}{w}.
\end{equation}

Applying the NLOC in \eqref{NLOC} for $\phi$, we obtain:

\begin{equation}
\begin{aligned}
\frac{\partial^2 \phi}{\partial z^2} &\approx \frac{-H_1 \phi(w, r) - H_2 \phi(0, r)}{w}\\
&\approx -\frac{H_1 + H_2}{w} \phi(r),\\
\end{aligned}
\end{equation}

assuming that \( \phi \) does not significantly vary along \( z \), which is valid under $w \rightarrow 0$. Substituting this into the governing equation, the resulting PDE simplifies to:

\begin{equation}
\frac{\partial^2 \phi}{\partial r^2} + \frac{1}{r} \frac{\partial \phi}{\partial r} - \left[ \frac{v^2}{4\alpha^2} + \frac{H_1 + H_2}{w} \right] \phi = -\frac{q'}{\alpha} \frac{\delta(r)}{r},
\end{equation}

which is one used by Rosenthal for the 2D model. This demonstrates that Rosenthal’s model naturally emerges from our generalized formulation under the thin-width assumption $w \rightarrow 0$. However, for finite values of \( w \), the Rosenthal assumption that \( \phi \) remains uniform along \( z \) no longer holds. Consequently, a more comprehensive treatment, such as the one provided by our model, is necessary to capture the cooling effects in a broader range of applications where \( w \) is non-negligible.

\subsection{The Limit of \( w \to \infty \): Recovering Rosenthal’s 3D Solution}\label{Rosenthal 3D}

To further validate our model, we examine the behavior of the temperature field in the limit \( w \to \infty \). Instead of taking this limit in the solution itself, a more systematic approach is to revisit the BVP.

To facilitate this analysis, we symmetrize the problem by shifting the coordinate system to the midpoint of the thin dimension \(z\), i.e., \( z \to z - \frac{w}{2} \). Under this transformation, the governing PDE is unchanged, while the boundary conditions can be rewritten as:

\begin{equation}
\left. T \right|_{(z = -w/2)} = e^{-H_1 w/2}, \quad \quad
\left. T \right|_{(z = w/2)} = e^{-H_2 w/2}.
\end{equation}

Since \( H_1 \) and \( H_2 \) are finite, the exponential terms decay rapidly as \( w \to \infty \), hence in this limit, B.C. are:

\begin{equation}
\left. T \right|_{z = -w/2} \approx 0, \quad \quad
\left. T \right|_{z = w/2} \approx 0.
\end{equation}

Thus, for large \( w \), the system behaves as an open domain in \( z \), restoring spherical symmetry. The governing PDE can then be rewritten in spherical coordinates \((\mathcal{R}, \theta, \phi)\), where \(\mathcal{R}\) is the radial coordinate:

\begin{equation}
\frac{\partial^2 \phi}{\partial \mathcal{R}^2} + \frac{2}{\mathcal{R}} \frac{\partial \phi}{\partial \mathcal{R}} - \frac{v^2}{4\alpha^2} \phi = 0.
\end{equation}

The general solution to this equation is:

\begin{equation}
\phi = A \frac{e^{-\frac{v}{2 \alpha} \mathcal{R}}}{\mathcal{R}},
\end{equation}

where \( A \) is determined by B.C. near the heat source at \( \mathcal{R} \approx 0 \). This recovers the well-known Rosenthal 3D solution, which does not include cooling effects. Thus, we have established that our model naturally reduces to the 3D Rosenthal solution in the $w \rightarrow \infty$, further confirming its consistency.

%\subsection{Unifying Multiple Regimes}

%By analyzing these limits, we demonstrate that our model encompasses various known cases of Rosenthal’s model:

%\begin{itemize}
%    \item \textbf{2D Rosenthal Model with Cooling:} Setting \( w \to 0 \) while retaining \( H_1 \) and \( H_2 \), our model reduces to the 2D Rosenthal model with cooling effects.
%    \item \textbf{3D Rosenthal Model without Cooling:} Taking \( w \to \infty \) eliminates cooling effects and restores the classical 3D Rosenthal solution.
%    \item \textbf{2D Rosenthal Model without Cooling:} Setting \( H_1 = H_2 = 0 \) in the \( w \to 0 \) limit retrieves the standard 2D Rosenthal model with no cooling effects.
%\end{itemize}

\subsection{Transient Behavior: Beyond Steady-State Solutions}\label{Transient Behavior}
As a proof of concept, we demonstrate that Rosenthal’s models can be extended beyond the steady-state solution, where \( \partial \phi /\partial t \neq 0 \). To probe more the transient behavior of the heat profile, we can use a tool such as the LT, which is particularly well-suited for IVPs. In contrast, using the FT in \( t \) can only recover the steady-state component of the solution. We solve an explicit case study within thin width approximation in Appendix \ref{SpecialCase LT Rosenthal}. Furthermore, we examine another case study within our generalized model, as detailed in Appendix \ref{TransientOurModel}.

In the following analysis, we continue working with the conformal profile \( \phi \). We focus on 2D model with cooling, where the PDE (with no steady state assumption) is given by:

\begin{equation}
\begin{gathered}
\frac{\partial \phi}{\partial t} - \alpha \left\{ \frac{\partial^2 \phi}{\partial x^2} + \frac{\partial^2 \phi}{\partial y^2} - \left[\frac{v^2}{4 \alpha^2} + \frac{H_1 + H_2}{w}\right] \phi \right\}\\ 
= q' \delta(x - x') \delta(y - y').
\end{gathered}
\end{equation}

While this PDE appears more complex than the steady-state case (Rosenthal's 2D heat equation), we can simplify the problem by leveraging the fact that in co-moving frame, the Kernel's source term is time-independent. Here, we first adopt the FT.

\textit{FT Approach: Recovering the Steady-State Solution}---The transformed governing equation takes the form:

\begin{equation}
\begin{gathered}
\frac{\partial^2 \phi'}{\partial r^2} + \frac{1}{r} \frac{\partial \phi'}{\partial r} - \left[\frac{v^2}{4\alpha^2} + \frac{H_1 + H_2}{w} + i\frac{\omega}{\alpha}\right] \phi' \\
= -\frac{2 \pi q'}{\alpha r} \delta(r) \delta(\omega),
\end{gathered}
\end{equation}
where $\phi \xrightarrow{FT} \phi^{'}$ in $\omega$- space.

This equation closely resembles Rosenthal’s steady-state model, with the only difference being the additional frequency-dependent term $i\frac{\omega}{\alpha}$. 

The presence of $\delta(\omega)$ ensures that integrating over $\omega$ reduces the solution to its steady-state counterpart:

\begin{equation}
T(t,r) = e^{-\frac{vx}{2\alpha}} \frac{q'}{\alpha} K_0\left(r \sqrt{\frac{v^2}{4\alpha^2} + \frac{H_1 + H_2}{w}}\right).
\end{equation}

This result confirms that the steady-state 2D Rosenthal solution naturally emerges in the Fourier domain. However, a fundamental limitation arises: \textit{the FT does not capture transient behavior—it inherently assumes the system has already reached equilibrium}. To investigate how the system evolves from its initial state to steady-state, we must turn to the LT.

\textit{LT Approach: Capturing Transient Behavior}---Applying the LT instead of the FT yields a governing equation with an additional term that accounts for I.C.:

\begin{equation}
\begin{gathered}
\frac{\partial^2 \phi'}{\partial r^2} + \frac{1}{r} \frac{\partial \phi'}{\partial r} - \left[\frac{v^2}{4 \alpha^2} + \frac{H_1 + H_2}{w} + \frac{s}{\alpha}\right] \phi'\\
= -\frac{2 \pi q'}{\alpha r} \frac{1}{s} \delta(r) - \phi(r,0).
\end{gathered}
\end{equation}
where this time $\phi \xrightarrow{LT} \phi^{'}$ in $s$-space.

The Laplace-transformed solution $\phi^{'}$ not only contains the steady-state term but also incorporates a transient component that encodes the system’s evolution from \( \phi(r,0) \) to the steady-state regime. The solution for $\phi$ in the Laplace domain is:
\begin{widetext}
    
\begin{equation}
\begin{gathered}
\phi(t,r) =  \frac{q' e^{-\frac{vx}{2\alpha}}}{\alpha} K_0\left(r \sqrt{\frac{v^2}{4\alpha^2} + \frac{H_1 + H_2}{w}}\right)+ \int_{\epsilon - i\infty}^{\epsilon + i\infty} ds \frac{e^{s\bar{t}}}{2\pi i}   \int^{\infty }_{0} dr'\phi(r',0) K_0\left(|r-r'| \sqrt{\frac{v^2}{4\alpha^2} + \frac{H_1 + H_2}{w} + \frac{s}{\alpha}}\right).
\end{gathered}
\end{equation}
\end{widetext}
The first term recovers the steady-state 2D Rosenthal solution, while the second term captures the transient behavior, illustrating how the LT effectively characterizes the system’s transition from its initial state to its steady-state phase. Consequently, if analyzing crack formation requires deeper insight into the transient state of the system, our model provides a framework to capture these dynamics as well.

For a concrete example of how the transient solution evolves under specific I.C., refer to Appendix \ref{SpecialCase LT Rosenthal}, where we illustrate through a case study the transient response. Additionally, we direct the reader to the Appendix \ref{TransientOurModel},that explores transient behavior in our full 3D model, extending beyond the thin-width approximation.

\section{Conclusion and Future Outlook}\label{Conclusion}

In this work, we developed a comprehensive analytical framework for finite-geometry 3D heat transfer while systematically incorporating cooling effects. By treating heat loss as an energy flux B.C, we provided a natural way to model cooling mechanisms. Our framework accommodates Newton's Law of Cooling, Stefan-Boltzmann radiation, and a generalized phenomenological cooling B.C, ensuring flexibility in modeling diverse physical systems.

A key focus of this study was on the transient behavior of the temperature profile, which plays a critical role in understanding crack formation during welding and additive manufacturing processes. We emphasized the significance of the Laplace Transform in capturing time evolution, as it effectively describes how the system evolves from its initial state toward equilibrium. The ability of the Laplace Transform to resolve both transient and steady-state behavior makes it particularly well-suited for problems involving time-dependent sources, which is why it was the primary tool in our analysis.

We analytically solved for time-dependent heat sources, such as the on/off switch point source model, obtaining closed-form solutions in 1D for moving sources and in 2D/3D for static sources. A generalization to a moving heat source in 3D results in a one-dimensional temporal integral, which can be efficiently evaluated numerically.

Beyond point sources, we extended our analysis to spatially distributed sources, including Gaussian, Ellipsoidal, and Double Ellipsoidal profiles, generalizing previous works such as those by Nguyen and Eager \& Tsai. In 3D, for short-duration sources, we provided closed-form analytical solutions using two equivalent approaches: (1) Two Fourier Transforms + Laplace Transform and (2) Two Fourier Transforms + Fourier Series. We proved their equivalence and demonstrated that, for long-duration sources, a one-dimensional integral is required, which remains computationally efficient for numerical evaluation.

Furthermore, we derived an analytical closed-form solution for the 3+1 Heat Kernel using both approaches mentioned above. Additionally, we validated the equivalence of different solution methods: (a) a pure numerical implementation of the full model, (b) Approach (1), and (c) Approach (2). Our results confirm that while all three methods yield equivalent solutions, analytical approaches significantly reduce computational costs and experimental resource requirements while also providing deeper physical insight into the dynamics of heat transfer.

Finally, we demonstrated that our framework naturally recovers Rosenthal’s models in 2D with cooling and 3D without cooling as special limiting cases, reinforcing the generality and robustness of our approach.

For future work, this methodology can be extended to incorporate finite-width boundaries in 2D and 3D using Fourier Series, which provides a natural extension of our current formulation. Another promising direction is the adaptation of our model to curved geometries, such as spherical, cylindrical, or elliptical domains, making it applicable to scenarios where sample geometry plays a crucial role. Furthermore, more complex heat sources can be analyzed by expanding them in Fourier or Taylor series in both time and space, allowing for the modeling of arbitrary heat deposition profiles.

By developing these extensions, our framework can be further refined to analyze more realistic heat transfer scenarios, aiding in computational modeling and experimental validation of thermal processes in materials science, manufacturing, and engineering applications. These advancements align with ongoing efforts within the Integrated Computational Materials Engineering (ICME) program to enhance predictive modeling capabilities for advanced manufacturing.

As a next step, we plan to conduct targeted experimental investigations to validate and benchmark our generalized thermal modeling framework. These experiments will focus on directly comparing the predictive accuracy of our model against other Rosenthal-based models reported in the literature.  By evaluating model performance under realistic process conditions — including variable heat source profiles, finite boundaries, and complex cooling mechanisms — we aim to further assess the robustness, applicability, and limitations of our approach. 

 Additionally, integrating our analytical framework with physics-informed machine learning models could enhance predictive heat management strategies. By leveraging ML techniques trained on synthesized data as well as experimental data, we can develop data-driven models that optimize heat control in manufacturing processes, ultimately leading to improved material quality, reduced thermal stresses, and enhanced process efficiency. Such hybrid approaches could bridge the gap between theoretical modeling and real-world applications.

\begin{acknowledgments}
We thank the EWI team for many fruitful discussions.
\end{acknowledgments}

\section*{Author Declarations}
\subsection*{Competing Interests}
The authors have no conflicts to disclose.
% (AIP requires a COI statement even if none.) :contentReference[oaicite:3]{index=3}

\subsection*{Author Contributions}
F.A.: conceptualization, theory, methodology, analysis, writing—original draft.
A.K.: validation, software, writing—review \& editing.
L.M.: validation, software, writing—review \& editing.
% (CRediT-style contributions are required.) :contentReference[oaicite:5]{index=5}

\section*{Data Availability}
The data that support the findings of this study are available from the corresponding author upon reasonable request.

\appendix
\clearpage

\section{On/Off Point Source}\label{source Using LT}
% (unchanged; LT properties are covered by \cite{schiff1999laplace})

Handling time-dependent source terms poses significant mathematical challenges as time is a shared variable across all spatial integrands in \eqref{Convlution}. Physically, this complexity arises from causality and retardation effects, encoded in the Heat Kernel $G$, which govern how heat propagates over time.

To simplify the analysis, we focus on a source term modeled as a point source with a capturing the gradual increase or decrease in power over time:

\begin{equation}
\begin{gathered}
S(t,x,y,z) =  \Theta_{\text{switch}}(t) A(t) \delta(x - x_{0}) \delta(y - y_{0}) \delta(z - z_{0}),\\
\Theta_{\text{switch}}(t)\equiv  \Theta(t - t_{\text{off}})-\Theta(t_{\text{on}} - t)
\end{gathered}
\end{equation}

$\Theta_{\text{switch}}(t)$ ``turns on" the heat source at \( t_{\text{on}} \) and ``turns it off" at \( t_{\text{off}} \), mimicking real-world heat pulse scenarios. As we will see below, generalizing \( A(t) \) to any polynomial or exponential function is straightforward. 

we choose to commute the ``source" integral with the spectral integral as discussed in \ref{Heat Profile for different Source terms}. While doing so may simplify the integration, it can introduce additional poles in the contour integral in the $s$-complex plane, as also further discussed in Appendix \ref{Using LT OPTION1}.

\subsubsection{1D Case: Heat Propagation Along \( z \)}
To build intuition, we first consider a simplified 1D case in the \( z \)-direction with cooling boundary conditions.

\textit{Constant Power Input (Step Function)} --- We begin by assuming a switched heat source with no explicit time dependence in power. This reduces the problem to evaluating the following integral:

\begin{widetext}
\begin{equation}
\begin{aligned}
T(t, z) &= \int_{-\infty}^{t} dt' \Theta_{\text{switch}}(t') \mathcal{L}^{-1} \left\{ \tilde{G}(z, z_{0}; s) \right\} 
= \int_{t_{\text{on}}}^{t_{\text{off}}} dt' \mathcal{L}^{-1} \left\{ \tilde{G}(z, z_{0}; s) \right\},
\end{aligned}
\end{equation}

where \( \mathcal{L}^{-1} \) denotes the inverse Laplace transform (ILT). Setting \( t_{\text{on}} = 0 \) and \( t_{\text{off}} = t \), meaning the source remains active until the moment of observation, we obtain:

\begin{equation}
T(t, z) = \int_{0}^{t} dt' \mathcal{L}^{-1} \left\{ \tilde{G}(z, z_{0}; s) \right\} = \mathcal{L}^{-1} \left\{ \frac{\tilde{G}(z, z_{0}; s)}{s} \right\}.
\end{equation}

This result provides the temperature profile while the source is active. If we also wish to describe the system after \( t_{\text{off}} \), we exploit the linearity of the problem:

\begin{equation}
\begin{gathered}
T(t, z) = -\int_{-\infty}^{\infty} dt' H(t' - t_{\text{on}}) \mathcal{L}^{-1} \left\{ \tilde{G}(z, z_{0}; s) \right\} - \int_{-\infty}^{\infty} dt' H(t' - t_{\text{off}}) \mathcal{L}^{-1} \left\{ \tilde{G}(z, z_{0}; s) \right\} \\
= -\int_{t_{\text{on}}}^{\infty} dt' \mathcal{L}^{-1} \left\{ \tilde{G}(z, z_{0}; s) \right\} - \int_{t_{\text{off}}}^{\infty} dt' \mathcal{L}^{-1} \left\{ \tilde{G}(z, z_{0}; s) \right\} \\
= \mathcal{L}^{-1}\left\{ \frac{\tilde{G}(z, z_{0}; s)}{s} \Big|_{t' = t_{\text{on}}} \right\} - \mathcal{L}^{-1}\left\{ \frac{\tilde{G}(z, z_{0}; s)}{s} \Big|_{t' = t_{\text{off}}} \right\}.
\end{gathered}
\end{equation}
\end{widetext}

This approach reduces the double integral to a single effective contour integral, at the expense of introducing a simple pole at \( s = 0 \) in the ILT, as anticipated. This integral can be evaluated using the Residue theorem, as discussed in \ref{Using the Residue Theorem}. A closed-form series solution follows naturally and corresponds to \eqref{ZPartSeriesSolution}, where \( A(\beta_n) \rightarrow A(\beta_n)/s_n \) and the series has an additional pole at \( s_n = 0 \).

\textit{Linearly Growing Power Input} --- Now, we consider a heat source with a linearly increasing power input during the heating phase:
\begin{equation}
A(t) = c_{0} + c_{1} t.
\end{equation}

Using the LT property:
\begin{equation}
\mathcal{L} \{ t^n f(t) \} = (-1)^n \frac{d^n}{ds^n} F(s),
\end{equation}
where \( n \) is a positive integer, we obtain:

\begin{widetext}
\begin{equation}
\begin{aligned}
T(t, z) &= c_{0} \mathcal{L}^{-1}\left\{ \frac{\tilde{G}(z, z_{0}; s)}{s} \Big|_{t' = t_{\text{on}}} \right\} 
- \mathcal{L}^{-1}\left\{ \frac{\tilde{G}(z, z_{0}; s)}{s} \Big|_{t' = t_{\text{off}}} \right\} \\
&\quad + c_{1} \mathcal{L}^{-1}\left\{ \frac{d}{ds} \left( \frac{\tilde{G}(z, z_{0}; s)}{s} \Big|_{t' = t_{\text{on}}} \right) \right\} 
- \mathcal{L}^{-1}\left\{ \frac{d}{ds} \left( \frac{\tilde{G}(z, z_{0}; s)}{s} \Big|_{t' = t_{\text{off}}} \right) \right\}.
\end{aligned}
\end{equation}
\end{widetext}

From here, obtaining a series solution is straightforward.

This result illustrates the advantage of working in spectral space: a potentially complicated temporal integration is transformed into a more manageable algebraic form, simplifying the solution.

\subsubsection{Static Source in 2D and 3D}

Now, let's extend the analysis to the 2D and 3D cases with a stationary source:

\begin{widetext}
\begin{equation}\label{SwitchStatic Source in 2D and 3D}
\begin{aligned}
T(t, x,y,z) &= \int_{t_{\text{on}}}^{t_{\text{off}}} dt' A(t') G_{0 1D} (x-x_{0}, \bar{t})  G_{0 1D}(y-y_{0}, \bar{t}) \mathcal{L}^{-1} \{\tilde{G}(z,z_{0};s)\}.
\end{aligned}
\end{equation}
\end{widetext}

We consider the static source case (\( v \to 0 \)) as performing the time integral analytically after commutation becomes not straightforward for \( v \neq 0 \). However, for \( v = 0 \), explicit evaluation is possible in certain cases. If the power input is constant, \( A(t) = A_0 \), the time integral can be carried out using error functions $\text{Erf}$, which leads to a different pole structure that can still be handled using the Residue Theorem, as discussed earlier. If \( A(t) \) is a polynomial in \( t \), the integral remains analytically solvable following similar steps. In contrast, for \( v \neq 0 \), the integral does not admit a closed-form solution and must be evaluated numerically.

\section{Option 1. Commuting Integrals: Ellipsoidal and Gaussian Sources Example}\label{Using LT OPTION1}
Returning to the case of an ellipsoidal Gaussian source, we aim to demonstrate that the heat profile \( T(t,x,y,z) \) can be obtained by commuting the integrals rather than evaluating the source integral first. Both methods are discussed in \ref{Heat Profile for different Source terms}. However, as we will show, this approach is analytically more involved for this specific example.

\begin{widetext}
\begin{equation}
\begin{aligned}
T(t, x, y, z) &= \int_{-\infty}^{t} dt' T_{x}(\bar{t},x) T_{y}(\bar{t},y) G_{0 1D}(y, \bar{t}) \int_{0}^{w} dz' e^{-\frac{(z' - z_c)^2}{\sigma_{z}^2}} \int_{\tilde{\epsilon} - i \infty}^{\tilde{\epsilon} + i \infty} ds e^{s \bar{t}} \tilde{G}(z, z')
\end{aligned}
\label{eq:T_solution}
\end{equation}

There is nothing forbidden about commuting the integrals and integrating the source terms with the \(z'\) dependence in \(\tilde{G}(z, z')\), leading to the following expression:

\begin{equation}
\mathcal{J}(\bar{t}, z) = \int_{\tilde{\epsilon} - i \infty}^{\tilde{\epsilon} + i \infty} ds \int_{-\infty}^{\infty} dz' e^{-\frac{(z' - z_c)^2}{\sigma^2}} e^{s \bar{t}} \tilde{G}(z, z') = \frac{\sqrt{\pi} \sigma_z}{2} \left(J_{(+, +)} + J_{(+, -)} + J_{(-, +)} + J_{(-, -)}\right)
\label{eq:I_final}
\end{equation}

Each \( J_{(\pm,\pm)} \) is given by:
\begin{equation}
J_{(\pm, \pm)} = \int_{\tilde{\epsilon} - i \infty}^{\tilde{\epsilon} + i \infty} ds \frac{j_{(\pm, \pm)}}{D(\beta, w, H_1, H_2)}
\label{eq:I_signs}
\end{equation}

where the denominator is defined as:
\begin{equation}
D(\beta, w, H_1, H_2) = 2 \beta \left[ (H_1 + \beta) (H_2 + \beta) - (H_1 - \beta) (H_2 - \beta) e^{-2 w \beta} \right]
\label{eq:D_beta}
\end{equation}

The numerator \( j_{\pm,\pm} \) is more intricate, involving error functions \( \operatorname{Erf} \), as shown below:

\begin{equation}
\begin{gathered}
j_{(+, -)} = -e^{(-2 w + z + z_c) \beta + \frac{\beta^2 \sigma^2}{4}} (H_1 + \beta) (H_2 - \beta) 
\left[ \operatorname{Erf}\left(\frac{w - z_c}{\sigma} - \frac{\beta \sigma}{2}\right) + 
\operatorname{Erf}\left(\frac{z_c}{\sigma} + \frac{\beta \sigma}{2}\right) \right]\\
j_{(-, +)} = -e^{\frac{1}{4} \beta (-4 z - 4 z_c + \beta \sigma^2)} (H_1 - \beta) (H_2 + \beta) 
\left[ \operatorname{Erf}\left(\frac{z_c}{\sigma} - \frac{\beta \sigma}{2}\right) + 
\operatorname{Erf}\left(\frac{2 w - 2 z_c + \beta \sigma^2}{2 \sigma}\right) \right]\\
j_{(+, +)} = e^{\frac{1}{4} \beta (-4 z - 4 z_c + \beta \sigma^2)} (H_1 + \beta) (H_2 + \beta) 
\left[ e^{2 z_c \beta} \operatorname{Erf}\left(\frac{z}{\sigma} - \frac{z_c}{\sigma} - \frac{\beta \sigma}{2}\right) + 
e^{2 z_c \beta} \operatorname{Erf}\left(\frac{z_c}{\sigma} + \frac{\beta \sigma}{2}\right) \right. \notag \\
 \quad \left. + e^{2 z \beta} \left\{ \operatorname{Erf}\left(\frac{2 w - 2 z_c + \beta \sigma^2}{2 \sigma}\right) - 
\operatorname{Erf}\left(\frac{2 z - 2 z_c + \beta \sigma^2}{2 \sigma}\right) \right\} \right]\\
j_{(-, -)} = e^{\frac{1}{4} \beta (-8 w - 4 z - 4 z_c + \beta \sigma^2)} (H_1 - \beta) (-H_2 + \beta) 
\left[ -e^{2 z_c \beta} \operatorname{Erf}\left(\frac{w}{\sigma} - \frac{z_c}{\sigma} - \frac{\beta \sigma}{2}\right) + 
e^{2 z_c \beta} \operatorname{Erf}\left(\frac{z}{\sigma} - \frac{z_c}{\sigma} - \frac{\beta \sigma}{2}\right) \right. \notag \\
 \quad \left. - e^{2 z \beta} \left\{\operatorname{Erf}\left(\frac{z_c}{\sigma} - \frac{\beta \sigma}{2}\right) + 
\operatorname{Erf}\left(\frac{2 z - 2 z_c + \beta \sigma^2}{2 \sigma}\right) \right\} \right].
\end{gathered}
\end{equation}

Although this integral appears quite involved, it retains the same pole structure as the Heat Kernel itself. Hence, we can find use Reside theorem and have a closed form series solution.

Moreover, from a computational perspective, this integral can be decomposed into 16 smaller sub-integrals using the linearity property of integration, in the same manner that the point source integral was split into four. 
\end{widetext}
\section{Modeling Heat Loss as Boundary Conditions}\label{Cooling Boundary Conditions}
NLOC describes the rate of heat loss of a body proportional to the difference in temperature between the body and its surroundings \cite{incropera2011fundamentals}:
\begin{equation}
\frac{\partial T}{\partial t} = -h (T - T_{\infty})
\end{equation}
where \( h \) is the heat transfer coefficient, \( T \) is the temperature of the body, and \( T_{\infty} \) is the ambient temperature. On the other hand, when considering SB radiation, the heat loss includes a term proportional to the fourth power of the temperature $T$, reflecting radiative heat transfer:
\begin{equation}\label{SBC}
\frac{\partial T}{\partial t} = -\epsilon \bar{\sigma} (T^4 - T_{\infty}^4)
\end{equation}
where \( \epsilon \) is the emissivity of the surface and \( \bar{\sigma} \) is the SB constant. Moreover, a phenomenological homogeneous cooling term can be expressed as:
\begin{equation}\label{PBC}
\frac{\partial T}{\partial t} = - \sum_{n=1}^{n'}  \mathcal{C}_n(T^n - T_{\infty}^n)
\end{equation}
where \(n'\) is a positive integer and \( \mathcal{C}_n \) represents coefficients for different powers of \( n \),  enabling a general model to account for different cooling mechanisms simultaneously.

When considering SB radiation, the heat loss includes a term proportional to $T^4$ \cite{siegel2015thermal}.

Since we mean by cooling loss of energy flux to the environment it we choose to model cooling as B.C. For example in the setup we have in this article, we choose to apply \eqref{NLOC}, \eqref{PBC}, and \eqref{SBC} at a condition on $T$ at $z=0$ and $z=w$.

Since cooling represents energy flux loss to the environment, we impose it as a B.C. In the framework of this study, we apply equations \eqref{NLOC}, \eqref{PBC}, or \eqref{SBC} at $z=0$ and $z=w$.

%\section{Modeling Heat Sources in AM}
%As the source can move with a time-dependent velocity \( v(t) \), and the power input \( P(t) \) may also fluctuate over time, incorporating these temporal variations is crucial for a more accurate model. A straightforward method to begin capturing this time dependence is by introducing polynomial or sinusoidal terms in \( t \), which can be viewed as truncated of a corresponding Taylor or FS, respectively. 

\section{Integral Transformations and Their Inverses}\label{FS and LT}
For foundational LT and inversion via Bromwich contour we refer to \cite{schiff1999laplace}. Standard differential-equation techniques referenced in \cite{krantz2022differential}.

The LT of a function \( f(t) \) is defined as
\begin{equation}
\mathcal{L}\{f(t)\} = F(s) = \int_0^\infty e^{-st} f(t) \, dt.
\end{equation}
Here, \( s \) is a complex variable, and the integral is evaluated over \( t \in [0, \infty) \). The Inverse Laplace Transform (ILT) recovers \( f(t) \) from \( F(s) \) and is given by
\begin{equation}\label{ILT}
f(t) = \mathcal{L}^{-1}\{F(s)\} = \frac{1}{2\pi i} \lim_{T \to \infty} \int_{\epsilon - iT}^{\epsilon + iT} e^{st} F(s) \, ds,
\end{equation}
where \( \epsilon \) is chosen such that all singularities of \( F(s) \) lie to the left of \(s_R = \epsilon \), ensuring convergence.

The FT decomposes a function into its frequency components and is particularly useful for infinite-domain problems. It is defined as
\begin{equation}
\mathcal{F}\{f(x)\} = \hat{f}(k) = \int_{-\infty}^{\infty} f(x) e^{-ikx} \, dx.
\end{equation}
The Inverse Fourier Transform (IFT) reconstructs \( f(x) \) from \( \hat{f}(k) \) using

\begin{equation}
f(x) = \mathcal{F}^{-1}\{\hat{f}(k)\} = \frac{1}{2\pi} \int_{-\infty}^{\infty} \hat{f}(k) e^{ikx} \, dk.
\end{equation}		  
\section{Why Are Initial Conditions Important?}\label{IC Section}

Understanding I.C. is essential for analyzing transient thermal behavior, which plays a critical role in crack formation. Rapid temperature changes create thermal gradients that induce internal stresses, leading to material failure. This is particularly important in AM processes, where uneven cooling rates can cause defects such as warping, dimensional inaccuracies, and phase transformations. Accurately modeling I.C. helps predict and mitigate these issues.

\textit{powder bed fusion (PBF)}---A relevant example arises in PBF processes, where each newly fused layer inherits thermal effects from the previously processed layers. The I.C. for a newly deposited layer is non-trivial due to residual heat retention. For instance, the initial temperature distribution can be modeled as an exponentially decaying function with depth:

\[
T(t=0, x, y, z) = T_0 e^{-\lambda z} \exp\left(-\frac{x^2}{\hat{\sigma}_{x}} -\frac{y^2}{\hat{\sigma}_{y} }\right),
\]

where \( T_0 \) is the surface temperature of the top layer, \( \lambda \) represents the thermal penetration depth into the underlying layers, and \( \hat{\sigma}_{x} \), \( \hat{\sigma}_{y} \) control heat spread in the spatial dimensions. This I.C. directly affects heat propagation in the new layer, influencing cooling rates and residual stresses.

\textit{Laser Cladding}---Another complex scenario arises in laser cladding, where a laser fuses powdered or wire material onto a substrate to form a coating or enhance surface properties. In this process, the I.C. is critical because the base material retains residual heat from previous passes. The new material being deposited must bond properly while avoiding defects like cracks or incomplete fusion. The temperature distribution in the substrate varies depending on the time between cladding passes and the thickness of the deposited layer. A suitable I.C. in this case may be an exponentially decaying temperature profile along both depth and radial directions:

\[
T(t=0, x, y, z) = T_0 e^{-\lambda z} \exp\left(-\alpha r^2 \right),
\]

where \( r = \sqrt{x^2 + y^2} \) represents the radial distance from the laser center, and \( \alpha \) describes the radial decay of heat. Properly accounting for retained heat in the substrate and the evolving initial temperature distribution is essential for ensuring a high-quality bond and preventing defects. 

\section{Derivation of Our Model's Green Function}\label{Derive GF}

The Green's function \( \tilde{G} \) for \eqref{Eqz} satisfies the ODE:

\begin{equation}
\gamma \tilde{G} - \frac{\partial^2 \tilde{G}}{\partial z^2} = \delta(z - z').
\end{equation}

Let \( \phi_H \) be the homogeneous solution of this ODE:

\begin{equation}
\phi_H = A e^{\beta z} + C e^{-\beta z}.
\end{equation}

Thus, \( \tilde{G}(z,z') \) can be written in the general form:

\begin{equation}\label{GreenFunctionWFormula}
\begin{gathered}
\tilde{G}(z, z') = \frac{\phi_{H}^{+}(z_>) \phi_{H}^{-}(z_<)}{W \left[ \phi_{H}^{+}(x), \phi_{H}^{-}(x)\right]},\\
W \left[ \phi_H^{+}(x), \phi_H^{-}(x) \right] = \phi_H^{+} \frac{\partial \phi_H^{-}}{\partial x}  - \phi_H^{-} \frac{\partial \phi_H^{+}}{\partial x},
\end{gathered}
\end{equation}

where \( - \) (\( + \)) indicates the solution satisfying the left (right) B.C. To solve for \( \phi_{H} \), we consider

\begin{equation}
\begin{gathered}
\gamma \phi_H -\frac{\partial^2 \phi_H}{\partial z^2} = 0,\\
\end{gathered}
\end{equation}

So far, we have enforced the B.C. in \( x \) and \( y \) through a Fourier integral. Next, focusing on the NLOC B.C, we use (\ref{NLOC}). The functions \( \phi_H^{-}(z_<) \) and  \( \phi_H^{+}(z_>) \) are given by:

\begin{equation}\label{leftBC}
\begin{gathered}
\phi_H^{-}(z_<) = C \left[ \eta e^{\beta z_<} + e^{-\beta z_<} \right],\\
\phi_H^+(z_>) = E \left[ e^{\beta z_>} + \xi e^{-\beta z_>} \right],\\
\eta = \frac{\beta + H_1}{\beta - H_1},\quad\quad
\xi = \frac{\beta + H_2 }{\beta - H_2}  e^{2\beta w}.
\end{gathered}
\end{equation}

Substituting (\ref{leftBC}) into (\ref{GreenFunctionWFormula}), we obtain

\begin{equation}
\begin{aligned}
\tilde{G}(z, z') &= \frac{1}{2 \beta (\eta \xi - 1)} \bigg[ \eta e^{\beta (z_> + z_<)} + e^{\beta (z_> - z_<)} \\
&\quad + \eta \xi e^{-\beta (z_> - z_<)} + \xi e^{-\beta (z_> + z_<)} \bigg].
\end{aligned}
\end{equation}

Following a similar approach, the Green's function can be derived for the SB B.C. (\ref{SBC}) or polynomial B.C. (\ref{PBC}). 

\section{Poles and Singular structure}\label{Poles and Singular structure}
For contour/branch-cut handling and estimation lemma background, see \cite{schiff1999laplace,churchill2009complex}.
We aim to prove by elimination that the poles $s_n$ of the integrand in \eqref{Isintegral} must be negative, purely real, and infinitely countable, thereby validating the factorization of the $x,y,z$ dependencies in the Green function. We start with the denominator of the integrand:

\begin{equation} \label{dem}
\begin{gathered}
(s^{1/2}-H_1)(s^{1/2}-H_2)e^{-2 s^{1/2}w}\\
- (s^{1/2}+H_1)(s^{1/2}+H_2) = 0.
\end{gathered}
\end{equation}

\textit{Case 1: The Root at $s=0$}---The equation has a simple root at $s=0$. However, this does not correspond to a pole of the integration, as the numerator in \eqref{Isintegral} also vanishes due to a simple root, ensuring that the integrand remains well-defined at $s=0$. Thus, there is no singularity at this point.

To facilitate further analysis, we introduce the following redefinitions:
\begin{equation}
\begin{gathered}
\beta \equiv \sqrt{\frac{s}{H_1 H_2}}, \\
\chi \equiv \frac{H_1+H_2}{\sqrt{H_1 H_2}} > 0, \\
\gamma_{+} \equiv \beta^2 + \chi \beta + 1, \\
\gamma_{-} \equiv \beta^2 - \chi \beta + 1, \\
\zeta \equiv 2 w \sqrt{H_1 H_2} > 0.
\end{gathered}
\end{equation}
Rewriting Eq. \eqref{dem} in terms of these variables, we obtain:
\begin{equation}
\gamma_{-} e^{-\zeta \beta } - \gamma_{+} = 0.
\end{equation}

\textit{Case 2: Real $\beta$ (Positive $s$)}---For $\beta > 0$, we note that:
\begin{equation}
\gamma_{+} > \gamma_{-} \Rightarrow \gamma_{+} > e^{-\zeta \beta} \gamma_{-}.
\end{equation}
Thus, the equation cannot be satisfied for $\beta > 0$. Similarly, for $\beta < 0$, we have:
\begin{equation}
\gamma_{+} < \gamma_{-} \Rightarrow \gamma_{+} < e^{-\zeta \beta} \gamma_{-}.
\end{equation}
Again, the equation is not satisfied. Therefore, no real values of $\beta$ (positive $s$) can be solutions.

\textit{Case 3: Purely Imaginary $\beta$ (Negative $s_n$)}---Setting $\beta \rightarrow i \tilde{\beta}$, we rewrite the equation in terms of trigonometric functions:
\begin{equation}
(-\tilde{\beta}^2 - i \alpha \tilde{\beta} + 1) e^{- i \zeta  \tilde{\beta}} - (-\tilde{\beta}^2 + i \alpha \tilde{\beta} + 1) = 0.
\end{equation}
Splitting into real and imaginary parts yields
\begin{equation}
\begin{gathered}
    (\cos(\zeta \tilde{\beta}) -1) (1-\tilde{\beta}^2)=\alpha \tilde{\beta} \sin(\zeta \tilde{\beta}),\\
    \alpha \tilde{\beta} (1+ \cos (\zeta \tilde{\beta}))=(\tilde{\beta}^2-1) \sin(\zeta \tilde{\beta}).
\end{gathered}
\end{equation}
This results in the following transcendental condition:
\begin{equation}
\cos(\zeta \tilde{\beta})= \frac{(H_1^2 -\tilde{\beta}^2)(H_2^2 -\tilde{\beta}^2)-4 H_1 H_2 \tilde{\beta}^2}{(H_1^2 + \tilde{\beta}^2)(H_2^2 + \tilde{\beta}^2)} <1,   
\end{equation}
which admits infinitely many solutions. These roots $s_n$ are purely real, negative, and not evenly spaced.

\textit{Case 4: Complex $\beta$ (Complex $s_n$)}---Repeating the analysis from Case 2, one can verify that the equation can only hold if the real part of $s_n$ is zero. Hence, complex roots with nonzero real parts are eliminated.

Since we have ruled out all cases except for purely real, negative roots $s_n$, we conclude that the only poles of the integrand lie on the negative real $s_R$-axis in spectral space, see Fig.~\ref{poles}.

\section{Fourier Series Approach: Full Derivation}\label{FS Derivation}

To obtain the Green’s function \( G \) using the FS approach, we expand the solution's $z$-dependence in terms of discrete eigenfunctions in $z$ dimension. This method naturally accommodates finite-width constraints and converges to the FT in the limit \( w \to \infty \), as discussed in the main text.

The eigenfunctions \( \phi_n(z) \) are determined by the BVP stemmed from \eqref{Eqz}:

\begin{equation}
\frac{d^2 \phi_n}{dz^2} + \lambda_n^2 \phi_n = 0,
\end{equation}

where the eigenvalues \( \lambda_n \) satisfy

\begin{equation}
\tan \left( \lambda_n w \right) = \frac{(H_1 + H_2) \lambda_n}{-H_1 H_2 + \lambda_n^2}.
\end{equation}

The corresponding eigenfunctions take the form

\begin{equation}
\begin{aligned}
\phi_{n}(z) &= \lambda_n \cos \left[\bar{z} \lambda_n \right] + H_2 \sin \left[\bar{z} \lambda_n \right] \\
&= \cos(z \lambda_n) + \frac{H_1}{\lambda_n} \sin(z \lambda_n).
\end{aligned}
\end{equation}

We expand \( \tilde{T}(t, k_x, k_y, z) \) and \( \tilde{S}(t, k_x, k_y, z) \) in terms of these eigenfunctions:

\begin{equation}
\begin{aligned}
\tilde{T}(t, k_x, k_y, z) &= \sum_{n} \hat{T}_n(t, k_x, k_y) \phi_n(z),\\
\tilde{S}(t, k_x, k_y, z) &= \sum_{n} \hat{S}_n(t, k_x, k_y) \phi_n(z).
\end{aligned}
\end{equation}

Similarly, the Green's function and Dirac delta function are expanded as

\begin{equation}
\begin{aligned}
\tilde{G}(t, k_x, k_y, z, z') &= \sum_{n} \phi_n(z) \phi_n(z') \tilde{G}_n(t, k_x, k_y),\\
\delta(z - z') &= \sum_{n} \phi_n(z) \phi_n(z').
\end{aligned}
\end{equation}

After applying this expansion, the heat equation reduces to an effective first-order ODE in \( t \):

\begin{equation}
\frac{\partial \tilde{G}_n}{\partial t} + \left( k_x^2 + k_y^2 - i k_x v + \lambda_n^2 \right) \tilde{G}_n = \delta(\bar{t}).
\end{equation}

Solving this ODE through direct integration gives

\begin{equation}
\tilde{G}_n = \Theta(\bar{t}) e^{-\left( k_x^2 + k_y^2 - i k_x v + \lambda_n^2 \right) \bar{t}}.
\end{equation}

Finally, performing the IFT yields the Green’s function in the spatial domain:

\begin{widetext}
\begin{equation}
G(t, x, y, z, x', y', z') = \Theta(\bar{t}) \sum_{n} \phi_n(z) \phi_n(z') \int_{-\infty}^{\infty} \int_{-\infty}^{\infty} e^{- \left( k_x^2 + k_y^2 - i k_x v + \lambda_n^2 \right) t} e^{i k_x (x - x') + i k_y (y - y')} \, dk_x \, dk_y.
\end{equation}

which simplifies to

\begin{equation}
G(t, x, y, z, x', y', z') = G_{0 1D}(\bar{y}, \bar{t}) G_{v 1D}(\bar{x},\bar{t}) \Theta(\bar{t}) \sum_{n} \phi_n(z) \phi_n(z') e^{- \lambda_n^2 \bar{t}}.
\end{equation}
\end{widetext}
This result confirms that the FS approach provides a fully equivalent solution to the LT approach, up to an eigenfunction normalization factor. This normalization introduces an additional constraint on the FT, specifically related to the I.C. of the system. On the other hand, the LT method inherently incorporates I.C, while in the FS approach, they must be explicitly imposed through normalization. The primary trade-off is that ILT involves evaluating a complex integral, whereas FS yields a series solution, making it computationally efficient when convergence is fast.

\section{Formulas}\label{Formulas}

\begin{widetext}

\begin{equation}
\begin{gathered}
    \mathcal{B}_{n}(z) = A(s_n) \left\{e^{-i z_c \beta_n} (i H_1 + \beta_n) 
    \left[ \operatorname{Erf}\left(a_{n}^{-}[z_c]\right) + \operatorname{Erf}\left(a_{n}^{+}[\tilde{z}]\right) \right] 
    + e^{i z_c \beta_n} (-i H_1 + \beta_n) 
    \left[ \operatorname{Erf}\left(a_{n}^{-}[\tilde{z}]\right) + \operatorname{Erf}\left(a_{n}^{+}[z_c]\right) \right]\right\}, \\
\end{gathered}
\end{equation}
\begin{equation}
\begin{gathered}
    \mathcal{C}_{n}(z) = A(s_n)  \left\{ e^{i \bar{z}_{c} \beta_n} (-i H_2 + \beta_n) 
    \left[ \operatorname{Erf}\left(-a_{n}^{+}[\bar{z}_{c}]\right) + \operatorname{Erf}\left(a_{n}^{+}[\tilde{z}]\right) \right] 
    + e^{-i \bar{z}_{c} \beta_n} (i H_2 + \beta_n) 
    \left[ \operatorname{Erf}\left(-a_{n}^{-}[\bar{z}_{c}]\right) + \operatorname{Erf}\left(a_{n}^{-}[\tilde{z}]\right) \right]\right\}, 
\end{gathered}
\end{equation}
\begin{equation}
\begin{gathered}
    a_{n}^{\pm}(z) \equiv \frac{1}{\sqrt{2}} \left(\frac{z}{\sigma_z} \pm i \beta_n \sigma_z\right), \\
    \tilde{z}\equiv z-z_c.
\end{gathered}
\end{equation}

\end{widetext}

\section{Illustration of Transient Behavior in 2D Model: A Case Study}\label{SpecialCase LT Rosenthal}

To better illustrate the role of transient dynamics in 2D model, we analyze a specific example where the I.C. is explicitly incorporated. As previously shown, the solution to the IVP consists of two terms: 

\begin{itemize}
    \item The first term, with appropriate handling, recovers the steady-state Rosenthal solution.
    \item The second term encodes information about the transient response, capturing the system’s evolution from its I.C. to steady-state.
\end{itemize}

To explore this transient behavior, we consider a physically reasonable initial condition:

\begin{equation}
\phi(r',0) = e^{-a r'}, \quad \text{where} \quad a > 0.
\end{equation}

Accordingly, the transient solution takes the form:
\begin{widetext}

\begin{equation}
\begin{gathered}
\phi_{\text{transient}}(r, t) = \frac{e^{-ar}}{2\pi i} \lim_{T \to \infty} \int_{\epsilon - iT}^{\epsilon + iT} ds \, e^{st}  
\left[ \int_0^r e^{a u} K_0(u B(s)) \, du + \frac{\ln\left(\frac{a + \sqrt{a^2 + B(s)^2}}{B(s)}\right)}{\sqrt{a^2 + B(s)^2}} \right],\\
B(s) = \sqrt{\frac{v^2}{4\alpha^2} + \frac{H_1 + H_2}{w} + \frac{s}{\alpha}}.
\end{gathered}
\end{equation}
\end{widetext}

For simplicity, consider the case where \( r \approx 0 \). In this regime, the first integral vanishes, and the I.C. modifies the pole structure, introducing explicit time dependence into the system.

The transient component reflects the system’s memory of its initial state. A fundamental question in heat conduction is determining the characteristic timescale over which the system transitions from its transient state to steady-state. This transition defines the point at which the influence of the initial condition is effectively erased, leaving only the steady-state solution dictated by the continuous heat source.

A more intense or time-dependent heat source would result in richer transient behavior. However, even with this simple example, using LT captures key physical features expected in practical applications such as welding, where heat propagation evolves over time before reaching a steady-state distribution.

This example highlights a key distinction between the Fourier and LT approaches:

\begin{itemize}
\item \textbf{FT:} Highlights intrinsic frequency modes, ideal for steady-state analysis. For a stationary source with time-translation symmetry, it retrieves steady-state behavior but omits transient evolution.
\item \textbf{LT:} Captures both transient and steady-state behavior, making it effective for analyzing I.Cs. and system relaxation timescales.
\end{itemize}

By incorporating the LT, we extend beyond Rosenthal’s steady-state analysis even in 2D and provide a more complete description of heat conduction. This approach allows us to investigate how the system evolves over time, offering insights into transient dynamics that are essential in applications such as welding and AM.

\section{Transient-State Analysis in Our 3D Model}\label{TransientOurModel}

In this appendix, we analyze the transient behavior of our full 3D model with cooling effects. Unlike steady-state solutions, which describe the long-term thermal distribution, transient solutions capture the system’s evolution from an initial condition to equilibrium. Understanding this transition is crucial in applications such as welding, where rapid thermal variations affect material properties.

In co-moving frame, the PDE for the $\phi$'s Kernel is
\begin{equation}
\begin{gathered}
\frac{\partial \phi}{\partial t} - \alpha \left\{ \frac{\partial^2 \phi}{\partial x^2} + \frac{\partial^2 \phi}{\partial y^2} + \frac{\partial^2 \phi}{\partial z^2} - \frac{v^2}{4 \alpha^2} \phi \right\}\\ = q' \delta(x - x') \delta(y - y') \delta(z - z').
\end{gathered}
\end{equation}

To solve this PDE, we consider two approaches: FT and separation of variables.

\textit{ FT}---Applying a FT in the \( x \)-\( y \) plane,
$\phi(x, y, z) \rightarrow \hat{\phi}(k_x, k_y, z)$, reduces the PDE to:

\begin{equation}
\begin{gathered}
-\bar{k}^2 \hat{\phi} + \frac{\partial^2 \hat{\phi}}{\partial z^2} = A.\\
\hat{\phi}(k_x, k_y, z) = C_{+} e^{\bar{k} z} + C_{-} e^{-\bar{k} z} - \frac{A}{\bar{k}^2}.
\end{gathered}
\end{equation}

where:
\begin{equation}
\begin{gathered}
\bar{k}^2 \equiv k_x^2 + k_y^2 + \frac{v^2}{4\alpha^2},\\ A \equiv -e^{\frac{v x'}{2\alpha}} \frac{q'}{\alpha} e^{-i(k_x x' + k_y y')},
\end{gathered}
\end{equation}

For the special case where \(H_1 = H_2 = h\), the coefficients simplify to:
\begin{equation}
\begin{gathered}
C_{\pm} = \frac{A h}{\bar{k}^2 \left[(h \pm \bar{k}) e^{\pm\bar{k} w} + h \mp \bar{k}\right]}\\
\end{gathered}
\end{equation}

\begin{widetext}

Reconstructing \(\phi(x, y, z)\) in real space via IFT:

\begin{equation}
\phi(x, y, z) = \frac{1}{(2\pi)^2} \int_{-\infty}^\infty \int_{-\infty}^\infty \hat{\phi}(k_x, k_y, z) e^{i(k_x x + k_y y)} \, dk_x \, dk_y.
\end{equation}

The temperature profile consists of two contributions: 

1. Free solution without boundary effects:
\begin{equation}
T_{\text{free}}(x, y) = \frac{q'}{2\pi \alpha} e^{\frac{-v(x-x')}{2 \alpha}} K_0\left(\frac{v}{2\alpha} \sqrt{(x - x')^2 + (y - y')^2}\right),
\end{equation}

2. Cooling contribution with B.C.:
\begin{equation}
T_{\text{cool}}(r,z) = \frac{q'}{\alpha} \frac{e^{\frac{-v \Delta x}{2\alpha}} }{(2\pi)^2} \int_{0}^\infty   dk  \, \frac{2 \sin{k r}}{  k^2 + \frac{v^2}{4 \alpha^2}} \left\{
\frac{H \left[e^{-\bar{k} z} + e^{-\bar{k}(w - z)}\right]}{(H - \bar{k}) e^{-\bar{k} w} + H + \bar{k}}
\right\}.
\end{equation}
\end{widetext}
\textit{ Separation of Variables}---Since solving the radial integral in Fourier space seems to be complicated analytically, we also consider solving the PDE using separation of variables. This method is particularly useful when modeling the heat source as a B.C. at \( r=0 \), similar to the original Rosenthal approach. We assume a solution of the form:

\begin{equation}
\phi(r, z) = \sum_{n} R_{n}(r) Z_{n}(z),
\end{equation}

which leads to the separated equations:

\begin{equation}
\frac{d^2 R_n}{d r^2} + \frac{1}{r} \frac{d R_n}{d r} - \left(\lambda_{n}^{2} + \frac{v^2}{4\alpha^2}\right) R_n(r) = 0,
\end{equation}

\begin{equation}
\frac{d^2 Z_n}{d z^2} + \lambda_{n}^{2} Z_n = 0.
\end{equation}

The solutions are:

\begin{equation}
R_n(r) = C K_0\left(\sqrt{\lambda_{n}^{2} + \frac{v^2}{4\alpha^2}} r\right),
\end{equation}

\begin{equation}
Z_n(z) = A_n \left(\cos{\lambda_n z} + B \sin{\lambda_n z}\right).
\end{equation}

Applying the NLOC B.C, this eigenvalues are given by:

\begin{equation}
\frac{\lambda_n}{H_2} \sin(\lambda_n w) -\frac{H_1}{H_2} \cos(\lambda_n w) = \cos(\lambda_n w) + \frac{ H_1}{\lambda_n}\sin(\lambda_n w).
\end{equation}

Finally, the full transient-state solution is:
\begin{widetext}
\begin{equation}
\phi= \sum_n A_n \left(\cos{\lambda_n z'} + \frac{H_1}{\lambda_n} \sin{\lambda_n z'}\right)\left( \cos{\lambda_n z} + \frac{H_1}{\lambda_n} \sin{\lambda_n z} \right)K_0\left(\sqrt{\lambda_{n}^{2} + \frac{v^2}{4\alpha^2}} r\right),
\end{equation}
\end{widetext}
where \( A_n \) is determined by the source condition at \( r=0 \).

Each approach offers valuable insights into heat propagation and cooling process. These solutions extend the steady-state analysis by capturing the full temporal evolution of the system, which is crucial in applications like welding and thermal modeling especially while examining crack formation.

\bibliography{main}   % <- NEW .bib name
\end{document}